\documentclass[sigconf,nonacm]{acmart}

\def\BibTeX{{\rm B\kern-.05em{\sc i\kern-.025em b}\kern-.08em
    T\kern-.1667em\lower.7ex\hbox{E}\kern-.125emX}}

\usepackage{booktabs} 
\usepackage{graphicx}
\usepackage{todonotes}
\usepackage{url}
\usepackage{caption}
\usepackage{multicol}
\usepackage{listings}
\usepackage{multirow}

\usepackage{dblfloatfix}

\usepackage{amsmath}
\usepackage{bm}
\begingroup\expandafter\expandafter\expandafter\endgroup
\expandafter\ifx\csname IncludeInRelease\endcsname\relax
\usepackage{fixltx2e}
\fi

\usepackage{nicematrix}
\usepackage{tikz}
\usepackage{enumitem}

\usepackage{color, colortbl}
\usepackage{rotating}

 \newcommand{\name}{$\text{Nextia}_{\text{JD}}$}

\makeatletter
\newsavebox{\measure@tikzpicture}
\NewEnviron{scaletikzpicturetowidth}[1]{%
	\def\tikz@width{#1}%
	\begin{lrbox}{\measure@tikzpicture}%
		\BODY
	\end{lrbox}%
	\pgfmathparse{#1/\wd\measure@tikzpicture}%
	\BODY
}
\makeatother

\tikzset{
	accepting/.style={
		append after command={
			\pgfextra{
				\begin{pgfinterruptpath}
					\draw [draw](\tikzlastnode) circle[radius=1em];
				\end{pgfinterruptpath}
			}
		}
	},
}
\usetikzlibrary{shapes}
\usetikzlibrary{arrows}
\usetikzlibrary{calc}
\usetikzlibrary{automata}
\usetikzlibrary{fit}                                    
\usetikzlibrary{backgrounds}    
\usetikzlibrary{decorations.pathreplacing}
\pgfdeclarelayer{bg}    
\pgfsetlayers{bg,main}  

\definecolor{d1}{gray}{0.95}
\definecolor{d2}{gray}{0.80}
\definecolor{d3}{gray}{0.65}
\definecolor{d4}{gray}{0.50}
\definecolor{d5}{gray}{0.35}

\usepackage{booktabs,subcaption,amsfonts,dcolumn}

\newcommand{\app}{\raise.17ex\hbox{$\scriptstyle\sim$}}

\setcopyright{none}

\acmDOI{}

\begin{document}
\title{Scalable Data Discovery Using Profiles}

\author{Javier Flores, Sergi Nadal, Oscar Romero}
\affiliation{
  \institution{Universitat Polit\`{e}cnica de Catalunya}
  \city{Barcelona}
  \state{Spain}
}
\email{jflores|snadal|oromero@essi.upc.edu}

\renewcommand{\shortauthors}{}

\begin{abstract}
We study the problem of discovering joinable datasets at scale. This is, how to automatically discover pairs of attributes in a massive collection of independent, heterogeneous datasets that can be joined. Exact (e.g., based on distinct values) and hash-based (e.g., based on \textit{locality-sensitive hashing}) techniques require indexing the entire dataset, which is unattainable at scale. To overcome this issue, we approach the problem from a learning perspective relying on profiles. These are succinct representations that capture the underlying characteristics of the schemata and data values of datasets, which can be efficiently extracted in a distributed and parallel fashion. Profiles are then compared, to predict the quality of a join operation among a pair of attributes from different datasets. In contrast to the state-of-the-art, we define a novel notion of join quality that relies on a metric considering both the containment and cardinality proportions between candidate attributes. We implement our approach in a system called \name,~and present extensive experiments to show the predictive performance and computational efficiency of our method. Our experiments show that \name~obtains similar predictive performance to that of hash-based methods, yet we are able to scale-up to larger volumes of data. Also, \name~generates a considerably less amount of false positives, which is a desirable feature at scale.

\end{abstract}

%
%


\maketitle

\section{Introduction}

Data discovery requires to identify interesting or relevant datasets that enable informed data analysis \cite{DBLP:conf/icde/FernandezMQEIMO18,DBLP:conf/icde/BogatuFP020}. Discovery and integration of datasets is nowadays a largely manual and arduous task that consumes up to 80\% of a data scientists' time \cite{DBLP:journals/debu/StonebrakerI18}. This only gets aggravated by the proliferation of large repositories of heterogeneous data, such as \textit{data lakes} \cite{DBLP:journals/pvldb/NargesianZMPA19} or open data-related initiatives \cite{DBLP:journals/debu/MillerNZCPA18}. For simplicity, from now on, we will refer to all such repositories as data lakes, which are valuable to contextualize and enrich third party data. However, data lakes commonly suffer from a lack of proper organization and structure \cite{DBLP:conf/sigmod/NargesianPZBM20}. Due to the unprecedented web-scale volumes of heterogeneous data sources, manual data discovery becomes an unfeasible task that calls for automation \cite{DBLP:conf/pods/GolshanHMT17}. Hence, we focus on the very first task of data discovery: i.e., the problem of discovering joinable attributes among structured datasets in a data lake. 

As an illustrative example of the challenges we face, take the reference dataset ($D_{ref}$) depicted in Table \ref{tab:happinessExample}.
\begin{table}[h!]
	\centering
	\begin{tabular}{|c|c|c|}
		\hline
		\textbf{Country} & \textbf{Happiness score} & \textbf{Schengen} \\
		\hline
		Mexico                                    & 6.595           & N  \\
		\hline
		Spain                                     & 6.354           & Y  \\
		\hline
		United States                             & 6.892           & N  \\
		\hline
		France                                    & 6.592           & Y  \\
		\hline
		Germany                                   & 6.985           & Y  \\
		\hline
	\end{tabular}
	\caption{\label{tab:happinessExample} $D_{ref}$ -- Happiness score per country in 2019}
\end{table}%
Assume we have a collection of other datasets available in the same data lake such as those depicted in Table \ref{tab:table_example_joins}. In such setting, we aim at finding joinable combinations of pairs of attributes from different datasets. A first observation is that purely schema-based methods, such as LogMap \cite{DBLP:conf/semweb/Jimenez-RuizG11}, would fail to propose the combination $D_{ref}.Country = D_1.X$ due to the lack of embedded semantics in the schema of $D_1$. Thus, we must also take into account the 
data values.

Note, however, that checking only 
data values might result in proposing the combination $D_{ref}.Schengen = D_2.Discount$, which is definitely not relevant for analysis. Furthermore, given positive pairs (i.e., likely to be meaningfully joinable), such as $D_{ref}.Country $ $= D_1.X$ and $D_{ref}.Country = D_2.Country$, there should be a clear criteria to rank them (i.e., suggest which one is \emph{better}). Ranking is relevant in web-scale scenarios, where independent data files must be crossed. Most of the times, in such scenarios, the cardinality of such files is not excessively large, but their order (i.e., number of columns / attributes) tend to be. Consequently, current approaches tend to propose too many joinable pairs of attributes, which is overwhelming for the end-user validating them. Rank suggestions is therefore highly advisable.
Finally, it would also be desirable that the attribute $D_3.Nation$ is suggested as joinable with $D_{ref}.Country$, even if it requires to apply some transformation. This case often holds when crossing independently created files.

Overall, the previous discussion can be summarized with the following questions, which serve as driver to motivate our work on large-scale data lake contexts:
\begin{itemize}[leftmargin=*]
	\item How to automatically detect candidate pairs of joinable attributes from large collections of heterogeneous, independent datasets?
	\item Can such join candidates be ranked according to the quality of the resulting join?

\end{itemize}

\begin{table}[!b]
	\centering
	\begin{tabular}{|c|c|c|}
		\hline
		\multicolumn{3}{|c|}{\textbf{Search accuracy}} \\
		\multicolumn{3}{|c|}{{\hspace{-3.5em} Exact\dotfill Approximate \hspace{-5em} \hfill}}                                                                                                                                                                              \\ \hline
		\begin{tabular}[c]{@{}c@{}}Comp. by value\\ \cite{DBLP:journals/pvldb/DengKMS17,DBLP:conf/sigmod/ZhuDNM19, DBLP:journals/tods/XiaoWLYW11}\end{tabular} & \begin{tabular}[c]{@{}c@{}}Comp. by hash\\ \cite{DBLP:conf/sequences/Broder97,DBLP:conf/icde/FernandezMNM19,DBLP:conf/icde/YangZZH19,DBLP:journals/pvldb/ZhuNPM16}\end{tabular} & \begin{tabular}[c]{@{}c@{}}Comp. by profile\\ \cite{DBLP:journals/debu/ChenGHTD18,DBLP:journals/ml/DoanDH03,DBLP:conf/esws/KejriwalM15a}\end{tabular} \\ \hline
		\multicolumn{3}{|c|}{{\hspace{-4.5em} Expensive\dotfill Efficient \hspace{-4.8em} \hfill }}                                                                                                                                                          \\
		\multicolumn{3}{|c|}{\textbf{Algorithmic complexity}} \\ \hline
	\end{tabular}%
	\captionof{table}{Overview of approaches by technique. These are further arranged according to the kind of search and their algorithmic complexity} \label{tab:relatedWork}
\end{table}
\begin{table*}[t!]
	\begin{center}
		\begin{subtable}{.25\linewidth}
			\centering
			\caption{\label{tab:populationExample} $D_1$ -- Population per country}
			\resizebox{\textwidth}{!}{
				\begin{tabular}{|c|c|c|}
					\hline
					\textbf{X} & \textbf{Y} & \textbf{Z} \\
					\hline
					Spain & 47M & 2020 \\
					\hline
					United States & 330M & 2020 \\
					\hline
					Mexico & 123M & 2020 \\
					\hline
					Germany & 83M & 2020 \\
					\hline
				\end{tabular}
			}
		\end{subtable}
		\begin{subtable}{.01\linewidth}
		\end{subtable}
		\begin{subtable}{.35\linewidth}
			\caption{\label{tab:storesExample} $D_2$ -- Customer satisfaction per store location}
			\centering
			\resizebox{\textwidth}{!}{
				\begin{tabular}{|c|c|c|c|}
					\hline
					\textbf{Country} & \textbf{Location} & \textbf{Discount} & \textbf{ \begin{tabular}[c]{@{}>{}c@{}}Customer \\satisfaction \end{tabular} } \\
					\hline
					United States & New York & Y & 7.7 \\
					\hline
					United States & Chicago & N & 8.5 \\
					\hline
					United States & Seattle & N & 8  \\
					\hline
					United States & Houston & Y & 7.7 \\
					\hline
				\end{tabular}
			}
		\end{subtable}%
		\begin{subtable}{.01\linewidth}
			\hspace*{\fill}
		\end{subtable}%
		\begin{subtable}{.375\linewidth}
			\centering
			\caption{\label{tab:expectancyExample} $D_3$ -- Average life expectancy}
			\resizebox{\textwidth}{!}{
				\begin{tabular}{|c|c|c|}
					\hline
					\textbf{Nation} & \textbf{ \begin{tabular}[c]{@{}>{}c@{}}Life expectancy \\(Women) \end{tabular} } & \textbf{ \begin{tabular}[c]{@{}>{}c@{}}Life expectancy \\(Men) \end{tabular} } \\
					\hline
					MX & 77.8 & 72.1 \\
					\hline
					SP & 86.1 & 86.1 \\
					\hline
					CA & 82.2 & 72.3 \\
					\hline
					US & 81.4 & 76.3 \\
					\hline
					BR & 79.4 & 72 \\
					\hline
				\end{tabular}
			}
		\end{subtable}
	\end{center}
	\caption{ \label{tab:table_example_joins} Three additional datasets of a toy-example repository ($D_1$, $D_2$ and $D_3$)}
\end{table*}

The problem of finding joinable attributes among datasets (or tables) is nowadays a relevant research topic of major interest for the data management community \cite{DBLP:conf/sigmod/BalazinskaCAFKS20,DBLP:journals/sigmod/AbadiAABBBBCCDD19}. We distinguish three approaches: \textit{comparison by value}, \textit{comparison by hash} and \textit{comparison by profile}. Table \ref{tab:relatedWork}, depicts an overview of recent contributions. Comparison by value relies on auxiliary data structures such as inverted indices or dictionaries to minimize the lookup cost.
Alternatively, the comparison by hash approach expects that similar values will collision in the same bucket, also employing index structures for efficient threshold index.
These approaches suffer from two major drawbacks. First, given that the similarity measure employed is either containment or Jaccard distance, precision is highly affected since they produce a large number of false positives as we experimentally discuss later.
A second drawback is the construction and maintenance of such index structures, a task that becomes highly demanding in terms of computing resources on large-scale datasets. In fact, as discussed in Section \ref{sec:evaluation}, current approaches do not scale-up. Oppositely, comparison by profile methods leverage on profiles extracted from datasets and their attributes. These are compared to predict whether a given pair of attributes will join. Thus, they are unaffected by the drawbacks posed by similarity measures, as they rely on the detection of similarities or discrepancies between profiles. Furthermore, working with summaries instead of data values is much more efficient from a complexity point of view. Yet, despite the clear performance benefits of profile-based approaches, there is nowadays a large gap in the trade-off regarding the quality of their results, as we experimentally observe.

In this paper, we aim to cover the gap generated by the low predictive performance of profile-based methods, as well as the limited precision and scalability of hash-based systems over data lakes.
We, thus, propose a novel learning-based method based on profiles to discover joinable attributes for large-scale data lakes. Our assumptions apply for scenarios where data are usually denormalized and file formats embed tabular data (i.e., not nested). We rely on state-of-the-art relational data profiling techniques \cite{DBLP:journals/vldb/AbedjanGN15} to compute informative profiles for datasets. This task, which can be done offline, allows us to extract and model the underlying characteristics of attributes. Furthermore, 
the computation of meta-features can be parallelized over 
distributed computing frameworks (e.g., Apache Spark). Next, profiles are compared in order to predict their expected join quality. Oppositely to the related work, which is mostly focused on containment or Jaccard distance, we also take into account the cardinality proportion between two attributes as an indicator of a higher join quality.
The predictive model is based on random forest classifiers, which are highly expressive and robust to outliers and noise \cite{DBLP:journals/ml/Breiman01}. Additionally, such models can be trained and evaluated in a distributed fashion \cite{DBLP:journals/pvldb/PandaHBB09}, thus yielding a fully distributed end-to-end framework for data discovery. 
We show that our method is generalizable and that proposes a meaningful ranking of pairs of attributes based on the predicted join quality.

\medskip

\noindent\textbf{Contributions.} We summarize our contributions as follows:
\begin{itemize}[leftmargin=*]
	\item We identify a set of profiling meta-features that allow to capture and model relevant characteristics of datasets and their attributes to predict their joinability.
	\item We introduce a qualitative metric for join quality, which considers both containment and the cardinality proportion between the candidate attributes.
	\item We learn a model based on random forest classifiers to efficiently rank candidate pairs of joinable attributes.
    \item We show that our approach is scalable and outperforms current state-of-the-art data discovery systems, while yielding higher predictive performance results 
    than current profile-based solutions and similar quality ($F_1$-score) to hash-based ones. Yet, relevantly, our approach yields better precision than hash-based approaches by proposing less false positives.
\end{itemize}

\medskip

\medskip

\noindent\textbf{Outline.}
The rest of the paper is structured as follows. We discuss related work and introduce the formal background, respectively in Sections \ref{sec:relatedwork} and \ref{sec:background}. Next, in Section \ref{sec:approach} we present the details of our approach, which we extensively evaluate in Section \ref{sec:evaluation}. We finally conclude our paper and present future work in Section \ref{sec:future}. 
\section{Related Work}\label{sec:relatedwork}

As depicted in Figure \ref{tab:relatedWork}, there are several techniques to tackle the problem of discovering joinable attributes, each with different trade-offs. Here, we survey relevant works for each category. Tables \ref{tab:relatedWork_value}, \ref{tab:relatedWork_hash} and \ref{tab:relatedWork_profile} present an overview of each of them.

\medskip

\noindent\textbf{Comparison by value.}
SilkMoth \cite{DBLP:journals/pvldb/DengKMS17} proposes a method to generate signatures from a subset of attribute tokens. To select an optimal subset, it uses an heuristic. Such signatures are used in an inverted index to prune the search space. Then, a verification step is required on the remaining candidates to discard those that do not hold for a certain similarity measure. This approach supports edit distance and Jaccard coefficient as similarity measures. Furthermore it assumes all signatures fit in memory. JOSIE \cite{DBLP:conf/sigmod/ZhuDNM19} proposes to optimize the number of comparisons scanning only the required values. To accomplish that, tokens are extracted from each attribute to create a dictionary and an inverted index.
\begin{table}[!h]
	\centering
	\resizebox{0.475\textwidth}{!}{

		\label{tab:definitions}
		\begin{tabular}{|r|c|c|c|}
			\hline
			~ & \textbf{ SilkMoth \cite{DBLP:journals/pvldb/DengKMS17} } & \textbf{ JOSIE \cite{DBLP:conf/sigmod/ZhuDNM19} } & \textbf{ PPJoin \cite{DBLP:journals/tods/XiaoWLYW11} } \\
			\hline
			Data type   & String & String  & String   \\
			\hline
			\begin{tabular}[c]{@{}r@{}}Similarity\\ measure\end{tabular} & \begin{tabular}[c]{@{}c@{}}Edit dist.\\and Jaccard\end{tabular} & Jaccard & \begin{tabular}[c]{@{}c@{}}Edit, Hamming,\\Jaccard and cosine\end{tabular} \\
			\hline
			\begin{tabular}[c]{@{}r@{}}Data\\ structure\end{tabular} & Inverted index & \begin{tabular}[c]{@{}c@{}}Dictionaries and\\inverted index\end{tabular} & Inverted index                                                    \\
			\hline
			\begin{tabular}[c]{@{}r@{}}Search\\ accuracy\end{tabular} & Exact & Exact & Approximate  \\
			\hline
			\begin{tabular}[c]{@{}r@{}}Search by\\threshold\end{tabular} & Yes & Yes & Yes \\
			\hline
			\begin{tabular}[c]{@{}r@{}}Kind of\\filter\end{tabular} & \begin{tabular}[c]{@{}c@{}}Token\\signatures\end{tabular} & Prefix, position & \begin{tabular}[c]{@{}c@{}}Prefix, position,\\suffix\end{tabular} \\
			\hline
		\end{tabular}
	}
	\captionof{table}{Overview of comparison by value approaches} \label{tab:relatedWork_value}
\end{table}
A ranked list is built from the $k$ most relevant candidate tables with highest containment, where $k$ is the number of candidates required.
Thus, attributes ranked at the top will have a larger number of common values. PPJoin \cite{DBLP:journals/tods/XiaoWLYW11} performs a different optimization by using a prefix filtering principle to avoid computing similarity values for all possible data values. As a result, this reduces the number of comparisons and hence improve efficiency. However, an inverted index requires a large space in memory. This approach proposes a similarity measure which combines tokens and characters.

\medskip

\noindent\textbf{Comparison by hash.}
MinHash \cite{DBLP:conf/sequences/Broder97} uses the minwise hash function from the LSH collection, with a collision probability equal to the Jaccard similarity. This requires, for every value, to compute the MinHash signature $K$ times, where $K$'s magnitude is in the hundreds. This approach has a major limitation on performance, as well as a bias towards small sets introduced by the Jaccard similarity. To overcome this, under the observation that MinHash can be optimized providing a Jaccard threshold, LSH Ensemble \cite{DBLP:journals/pvldb/ZhuNPM16} proposes to use containment similarity and convert it to a Jaccard threshold. It focuses on finding attributes with a high containment similarity, that is, to cover as many values as possible. For efficient indexing, LSH Ensemble partitions the sets according to the set size. However, this introduces false positives, that is, attributes that do not fulfill the similarity threshold used. GB-KMV \cite{DBLP:conf/icde/YangZZH19} aims to reduce the number of false positives generated by LSH Ensemble. Further, it considers that additional information (e.g., attribute cardinalities and value frequencies) to offer better performance in estimating the containment similarity. Another approach that aims to tackle the drawbacks of MinHash is Lazo \cite{DBLP:conf/icde/FernandezMNM19}. Here, the Jaccard similarity is redefined to consider set cardinalities, which allows to estimate the containment similarity. Instead of computing $K$ times a hash function, Lazo implements the One Permutation Hashing (OPH) technique, hashing data values only once. This approach, however, also introduces false positives. In order to minimize them, an heuristic method based on cardinalities is introduced. A distinguishable feature of Lazo is that rehashing the entire dataset collection is not required when a new one is introduced.

\begin{table}[h!]
	\centering
	\resizebox{.475\textwidth}{!}{%
		\begin{tabular}{|r|c|c|c|c|}
			\hline
			\multicolumn{1}{|l|}{}                                       & \textbf{ MinHash \cite{DBLP:conf/sequences/Broder97} } & \textbf{ \begin{tabular}[c]{@{}c@{}}LSH\\ Ensemble \cite{DBLP:journals/pvldb/ZhuNPM16}\end{tabular} } & \textbf{ GB-KMV \cite{DBLP:conf/icde/YangZZH19} } & \textbf{ Lazo \cite{DBLP:conf/icde/FernandezMNM19} } \\ \hline
			Datatype                                                     & String           & String                & String          & String                                                            \\ \hline
			\begin{tabular}[c]{@{}r@{}}Similarity\\ measure\end{tabular} & Jaccard          & Containment           & Containment     & \begin{tabular}[c]{@{}c@{}}Jaccard and\\ Containment\end{tabular} \\ \hline
			\begin{tabular}[c]{@{}r@{}}Threshold\\ index\end{tabular}    & Yes              & Yes                   & Yes             & No                                                                \\ \hline
		\end{tabular}%
	}
	\captionof{table}{Overview of comparison by hash approaches} \label{tab:relatedWork_hash}
\end{table}

\medskip

\noindent\textbf{Comparison by profile.}
LSD \cite{DBLP:journals/ml/DoanDH03} proposes a multi-strategy learning approach to automatically find related attributes among XML files. It applies multiple learner modules, where each module exploits different kind of information, either from schema or data values. Such predictions are combined to weight each learner. LSD also exploits domain integrity constraints and user feedback. FlexMatcher \cite{DBLP:journals/debu/ChenGHTD18} extends LSD with more data types. A relevant aspect is that it considers pattern classifiers to filter data values. A limitation of this approach is that every time a discovery process is to be performed on a specific attribute, it requires to train new models providing a training sample of attributes that might join with the specific attribute. 
A different approach is that of \textit{Semi-supervised Instance Matching Using Boosted Classifiers} (SIMUBC) \cite{DBLP:conf/esws/KejriwalM15a}, which aims to detect pairs of attributes sharing common values. SIMUBC extracts 28 kinds of metadata from attributes such as tokens, phonetic values or representatives. Such metadata are used to train Random Forest and Multilayer Perceptron models to predict whether two attributes are join candidates. To improve performance, weights are assigned to each model to compute the final prediction. A limitation of this work is that it requires to train the models each time a data discovery process is started.

\begin{table}[h!]
	\centering
	\resizebox{.475\textwidth}{!}{%
		\begin{tabular}{|r|c|c|c|}
			\hline
			~                           & \textbf{LSD \cite{DBLP:journals/ml/DoanDH03} } & \textbf{ FlexMatcher \cite{DBLP:journals/debu/ChenGHTD18} } & \textbf{ SIMUBC \cite{DBLP:conf/esws/KejriwalM15a} }  \\
			\hline
			Data type                   & \begin{tabular}[c]{@{}c@{}}String and\\numeric\end{tabular}                 & \begin{tabular}[c]{@{}c@{}}String \\and numeric\end{tabular}                   & String                                                                                                                                                             \\
			\hline
			File format                 & XML                                                                         & CSV, JSON                                                                      & NA                                                                                                                                                                 \\
			\hline
			Technique~                  & Classification                                                              & Classification                                                                 & Classification                                                                                                                                                     \\
			\hline
			\begin{tabular}[c]{@{}r@{}}ML\\algorithm\end{tabular} & \begin{tabular}[c]{@{}c@{}}Naive Bayes and \\nearest-neighbor\end{tabular} & \begin{tabular}[c]{@{}c@{}}Random forest,\\nearest neighbors,\\ and logistic regression\end{tabular} &
			\begin{tabular}[c]{@{}c@{}}Random forest\\and multilayer\\perceptrons\end{tabular} \\
			\hline
			\begin{tabular}[c]{@{}r@{}}Accuracy \\search\end{tabular}  & Approximate & Approximate  & Approximate                                                                                                                                                        \\
			\hline
		\end{tabular}
	}
	\captionof{table}{Overview of comparison by profile approaches} \label{tab:relatedWork_profile}
\end{table}

\section{Background}\label{sec:background}

This section introduces the formal background and state our assumptions. 

\subsection{Data model}\label{sec:datamodel}

\noindent\textbf{Data repositories and datasets.}
A data repository $\mathcal{D}$ is a finite nonempty set of dataset names $\{ D_1, \ldots, D_m \}$, where each $D_i$ has a fixed arity $n_i$. Let $A$ be a set of attribute names, then each $D_i \in \mathcal{D}$ is associated to a tuple of attributes denoted by $att(D_i)$. Henceforth, we will assume that $\forall i,j : i \neq j  \rightarrow att(D_i) \cap att(D_j) = \emptyset$. 
We use $att(\mathcal{D})$ to refer to the set $\{ atts(D_1) \cup \ldots \cup atts(D_n) \}$. Then, let $V$ be a set of values, a tuple $t$ in $D_i$ is a function $t: att(D_i) \rightarrow V$. For any dataset $D_i$, $tuples(D_i)$ denotes the set of all tuples of $D_i$.

\medskip

\noindent\textbf{Conjunctive queries}.
We aim at discovering conjunctive queries (CQs), where each CQ is a expression of the form
\begin{equation*}
Q = \pi_{\overline{y}}(D_1 \times \ldots \times D_n) \textsc{ } | \bigwedge\limits_{i=1}^m P_i(\overline{z_i}) )
\end{equation*}
where $D_1, \ldots, D_n$ are distinct datasets; $P_1, \ldots, P_n$ are equi-join predicates respectively over $\overline{z_1}, \ldots \overline{z_n}$; and both $\bigcup_{i=1}^m \overline{z_i}$ and $\overline{y}$ are subsets of $\bigcup_{i=1}^n att(D_i)$. We may also refer to binary equi-join predicates as pairs of the form $p = \langle a_1,a_2 \rangle$, noting that $\langle a_1,a_2 \rangle = \langle a_2,a_1 \rangle$.
We will assume that CQs 
are evaluated under \textit{set semantics}, and refer the reader to \cite{DBLP:books/aw/AbiteboulHV95} for the formal aspects on their evaluation.
We use $Q(\mathcal{D})$ to denote the execution of a CQ $Q$ over a data repository $\mathcal{D}$, a function returning a set of tuples $T \subseteq \{ tuples(D_1) \times \ldots \times tuples(D_n) \}$ over $\overline{y}$.
Hereinafter we will also assume that all equi-join predicates $P_i$ only refer to string attributes. Note that we do not only check the attribute type defined in the dataset, but also the real datatype. As an example, numeric values such as phone numbers encoded as string values will still not be considered strings and therefore ruled out.

\begin{table*}[!b]
	\begin{center}
		\begin{subtable}{.275\linewidth}
			\centering
			\caption{\label{tab:tourismIncome} $D_{ref}$ -- Tourism income in Spain}
			\resizebox{\textwidth}{!}{
				\begin{tabular}{|c|c|c|}
					\hline
					\textbf{City} & \textbf{Seaside} & \textbf{Amount}  \\
					\hline
					Barcelona & Y & 350M \\
					\hline
					Girona & Y & 110M \\
					\hline
					Lleida & N & 75M \\
					\hline
					Tarragona & Y & 83M \\
					\hline
					\ldots & \ldots & \ldots \\
					\hline
				\end{tabular}
			}
		\end{subtable}
		\begin{subtable}{.01\linewidth}
		\end{subtable}
		\begin{subtable}{.4\linewidth}
			\caption{\label{tab:demographicEU} $D_1$ -- EU demographic data}
			\centering
			\resizebox{\textwidth}{!}{
				\begin{tabular}{|c|c|c|c|}
					\hline
					\textbf{Unit} & \textbf{Population} & \textbf{Avg. salary} & \textbf{Cost of living} \\
					\hline
					Antwerp & 1,120,000 & 44,000€ & 2,896€ \\
					\hline
					Barcelona & 1,620,343 & 31,000€ & 2,422€ \\
					\hline
					Berlin & 4,725,000 & 49,000€ & 2,737€  \\
					\hline
					Bristol & 1,157,937 & 30,000£ & 2,397£ \\
					\hline
					\ldots & \ldots & \ldots & \ldots \\
					\hline
				\end{tabular}
			}
		\end{subtable}%
		\begin{subtable}{.01\linewidth}
			\hspace*{\fill}
		\end{subtable}%
		\begin{subtable}{.3\linewidth}
			\centering
			\caption{\label{tab:demographicWorld} $D_2$ -- Worldwide demographic data}
			\resizebox{\textwidth}{!}{
				\begin{tabular}{|c|c|c|}
					\hline
					\textbf{Name} & \textbf{Country} & \textbf{ Population } \\
					\hline
					Barcelona & Spain & 1,620,343 \\
					\hline
					Canberra & Australia & 426,704 \\
					\hline
					Chicago & United States & 2,695,598 \\
					\hline
					Curitiba & Brasil & 1,908,359 \\
					\hline
					\ldots & \ldots & \ldots \\
					\hline
				\end{tabular}
			}
		\end{subtable}
	\end{center}
	\caption{ \label{tab:motivateJoinQuality} A reference dataset ($D_{ref}$) and two candidate datasets to be joined. $D_1$ is curated with extensive data at european level, while $D_2$ is curated at the worldwide level with less details}
\end{table*}

\subsection{Similarity measures}

We distinguish two different perspectives, yet complementary, to identify joinable attributes: based on \textit{the relationship between the candidate join attributes} and by means of \textit{quantifiable measures}. 

\medskip

\noindent\textbf{Measures based on attribute relationships.} 
\textit{Syntactic} and \textit{semantic} relationships for pairs of attributes refer, in a qualitative way, to the kind of form and meaning of their relationship. Syntactic relationships refer to the form of the data. Two attributes have a syntactic relationship when data values have the same form, and thus they share common values. Nevertheless, as occurs for the pair \textit{Schengen} (in Table~\ref{tab:happinessExample}) and \textit{Discount} (in Table~\ref{tab:storesExample}), values might have a different meaning. To this end, we consider semantic relationships, which consider the meaning of the values. Thus two attributes have a semantic relationship if data values refer to the same concept in a shared domain. In general, attributes with a semantic relationship also have a syntactic one. When this is not satisfied, as happens for the pair \textit{Country} (in Table~\ref{tab:happinessExample}) and \textit{Nation} (in Table~\ref{tab:expectancyExample}), we refer to this relationship as \textit{semantic non-syntactic}. 

\medskip

\noindent\textbf{Quantifiable measures.}
A quantifiable way to define that a pair of attributes share common values is by using a similarity measure indicating the degree of similarity between them. When proposing joinable attributes, only quantifying the similarity between two attributes without considering their relationship generates large numbers of false positives. However, due to the difficulty in identifying the kind of relationship between a pair of attributes, existing approaches propose joinable attributes exclusively based on quantifiable similarity measures. These are chosen based on the data type. Here, we focus on pairs of sets $A,B$ of string attributes. Two of the most commonly used measures are containment similarity ($C(A,B)$) and Jaccard similarity ($J(A,B)$), formalized as:

\medskip

\begin{minipage}{13em}
	\begin{equation*} \label{eq:containmentScore}
	C(A,B) = \frac{ |A \cap B|}{|A|}
	\end{equation*}
\end{minipage}
\begin{minipage}{0em}
	\begin{equation*} \label{eq:jaccardSimilarity}
	J(A,B) = \frac{ |A \cap B|}{|A \cup B|}
	\end{equation*}
\end{minipage}

\medskip

These metrics work on sets (i.e., well-defined collections of distinct values). Note that Jaccard similarity is symmetric, thus it can be biased towards smaller sets. Oppositely, containment similarity measures the relative size of the intersection of two sets over the size of one of them. Hence, such measure is asymmetric. In both cases, both measures range from 0 to 1. A similarity closer to 1 denotes that two attributes are more similar.

\subsection{Measuring the quality of a join} \label{subSection:qualiyJoin}

Unlike the state-of-the-art, which mainly uses containment and Jaccard similarities to decide the degree of joinability among pairs of attributes, we define a qualitative metric to measure the expected join quality.
We consider containment as a desirable metric to maximize. Yet, we make the observation that datasets on a data lake do not relate to each other as in a relational database. In such scenarios, it is common to find datasets with few data values in common. In order to exemplify this idea, let us consider the datasets depicted in Table \ref{tab:motivateJoinQuality}. In this example, the reference dataset $D_{ref}$ might be joined with any of the two candidate datasets $D_1$ (at the EU level) and $D_2$ (worldwide). Current approaches would propose both as positive pairs, since they yield the same containment. However, we aim at distinguishing the join quality between them and use their \emph{cardinality proportion} for that purpose. Let us consider the following cardinalities corresponding to the city attributes: |$D_{ref}$| = 8124, |$D_1$| = 54500 and |$D_2$| = 982921. We use the cardinality proportion as a measure to infer whether their data granularities are similar. In this sense, the third dataset is much larger than |$D_{ref}$| and yield a worse proportion and therefore we rank it worse. Importantly, we assume these datasets store independently generated events and such big difference in their cardinality most probably mean they embed different semantics or sit at different granularity levels. In general, such situations are a source of false positives for current solutions, specially, when considering small tables.

\medskip

\noindent\textbf{Join quality.}
We now formalize the metric for join quality as a rule-based measure combining both containment and cardinality proportion. To this end, we define a totally-ordered set of quality classes $S = \{ \texttt{None}, \texttt{Poor}, \texttt{Moderate}, \texttt{Good}, \texttt{High} \}$ as indicator of the quality of the resulting join. Indeed, we advocate it is not possible to define a binary class (i.e., either joinable or not), since we would not be able to rank the positive ones. 
As experienced with the state-of-the-art, in realistic large scenarios, a binary class yields a long list of results difficult to explore. Therefore, we propose the following multi-class join quality metric.

\begin{definition}
Let $A, B$ be sets of values, respectively the \textit{reference} and \textit{candidate} attributes. The join quality among $A$ and $B$ is defined by the expression
\[\small
Quality(A, B) = \normalsize
\begin{cases}
(4)\text{ }\texttt{High}, & C(A,B) \geq C_H \wedge \frac{|A|}{|B|} \geq K_H \\ 
(3)\text{ }\texttt{Good}, & C(A,B) \geq C_G \wedge \frac{|A|}{|B|} \geq K_G \\ 
(2)\text{ }\texttt{Moderate}, & C(A,B) \geq C_M \wedge \frac{|A|}{|B|} \geq K_M \\ 
(1)\text{ }\texttt{Poor}, & C(A,B) \geq C_P\\
(0)\text{ }\texttt{None}, & \text{otherwise}
\end{cases}
\]
\end{definition}

The rationale behind the quality metric is to constraint the candidate pairs of attributes to two thresholds per class: containment (i.e., $C_{i}$) and cardinality proportion (i.e., $S_{i}$). Precisely, we fix that for any pair of classes $S_i, S_j \in S$ where $S_i > S_j$, the containment proportion must be higher (i.e., $C_H > C_G > C_M > C_P$) and the cardinality proportion must be smaller ($K_H > K_G > K_M$). Intuitively, a larger containment and smaller cardinality proportion guarantees that the two attributes share common values and their cardinalities are alike. Consequently, most probably, they have a semantic relationship.
We consider the values $C_H = 3/4 = 0.75, C_G = 2/4 = 0.5, C_M = 1/4 = 0.25, C_P = 0.1$ for containment, and $K_H = 1/4 = 0.25, K_G = 1/8 = 0.125, K_M = 1/12 = 0.083$ for cardinality proportion. These have been empirically defined from our training set, yet, as we show in Section \ref{sec:evaluation} they are generalizable to other datasets.

Taking again the example from Table \ref{tab:motivateJoinQuality}, Table \ref{tab:qualities} shows the values corresponding to containment and cardinality proportion between the attributes about cities. Note that, although the containment is very high in both cases, the constraint on cardinality proportions assigns a \texttt{Poor} class to the combination $D_{ref},D_2$. Although both pairs are predicted as joinable, our goal is to rank in a higher position the pair $D_{ref},D_1$ denoting a more interesting join result. Importantly, the cardinality proportion constraint does not apply when the reference dataset is the one much larger. In such cases, the containment should be the leading factor, and thus determine the ranking. Realize that, due to set semantics, it is not possible for the reference attribute to be comparatively larger than the candidate one, and yield a high containment.
\begin{table}[h!]
	\centering
	\begin{tabular}{|c|c|c|c|}
		\hline
		\textbf{Pair} & \textbf{Cont.} & \textbf{Card. prop} & \textbf{Quality} \\
		\hline
		$D_{ref}.City,D_1.Unit$ & 0.8 & 0.149 & 3 \\
		\hline
		$D_{ref}.City,D_2.Name$ & 0.95 & 0.00826 & 1 \\
		\hline
	\end{tabular}
	\caption{\label{tab:qualities} Qualities for the datasets depicted in Table \ref{tab:motivateJoinQuality}}
\end{table}%

To showcase the benefit of considering the cardinality proportion consider the following extreme case, which is not uncommon on large-scale automated data discovery scenarios. Consider the two datasets depicted in Tables \ref{tab:cities} and \ref{tab:movies}, the former ($D_s$) listing the opening hours of stores and the latter ($D_m$) movies and their directors. Let us assume |$D_s$| = 3 and |$D_m$| is above a million movies. Current solutions, which exclusively consider containment, would qualify the pair $\langle D_s.Store, D_m.Movie \rangle$ pair as a high quality join, given the 2/3 containment (which would be even higher if we consider approximate joins). Yet, this is clearly a false positive. Considering the cardinality proportion, our quality metric would penalize its ranking and assign a \texttt{Poor} label to this candidate pair.

\begin{table}[h!]
	\centering
	\begin{tabular}{|c|c|c|}
		\hline
		\textbf{Store} & \textbf{Opening hour} & \textbf{Closing hour} \\
		\hline
		Chicago & 8am & 18pm  \\
		\hline
		Casablanca & 9:30am & 20pm  \\
		\hline
		Paris & 9am & 18pm  \\
		\hline
	\end{tabular}
	\caption{\label{tab:cities} $D_{s}$ -- Store schedules}
	
	\begin{tabular}{|c|c|}
		\hline
		\textbf{Movie} & \textbf{Director} \\
		\hline
		An American in Paris & George Gershwin  \\
		\hline
		Casablanca & Michael Curtiz  \\
		\hline
		Chicago & Rob Marshall \\
		\hline
		\ldots & \ldots \\
		\hline
	\end{tabular}
	\caption{\label{tab:movies} $D_{m}$ -- Movies and their directors}
\end{table}%

\subsection{Join discovery as a classification problem}\label{sec:statement}

\noindent\textbf{Profiles.}
A unary profile $P_u$ for an attribute $A$, referred as $P_u(A)$ is a set of meta-features $\{ m_1, \ldots, m_n \}$. Each $m_i$ is a summary or statistic about the structure or content of $A$ (e.g., number of distinct values). We also consider binary profiles, which are meta-features that denote characteristics of a relationship between pairs of attributes. Hence, we define a binary profile $P_b$ for a pair of attributes $A,B$, denoted $P_b(A,B)$, as a set of meta-features (e.g., Levenshtein distance from $A$ to $B$).

\medskip

\noindent\textbf{Predicting the join quality.}
Computing $Quality(A,B)$ for any pair of attributes $A,B$ is naturally unattainable at scale. Hence, as previously discussed, we tackle the join discovery problem as a classification problem. To this end, we aim to predict the degree of similarity of candidate pairs $\langle A,B \rangle$ from $A$'s and $B$'s profiles, which we will use as an indicator of joinability and predict its quality. We, thus, define the following function:

\begin{definition}\label{def:predict}
Let $C: C_1, \ldots, C_n$ be a totally-ordered set of classes, where for any $i > j$, $C_i$ indicates an equal or higher degree of similarity than $C_j$. We define a function $Predict(P_u(A),P_u(B),$ $P_b(A,B))$ such that for a pair of attributes $A,B$ yields a predicted similarity class $S_i \in S$ from their unary and binary profiles.
\end{definition}


\noindent\textbf{Problem statement.}
We now formalize the multi-class classification join discovery problem. The goal is to discover a ranking (i.e., a partially-ordered set) of equi-join predicates based on their predicted join quality. We distinguish two settings for the join discovery problem: \textit{discovery-by-attribute} and \textit{discovery-by-dataset}. The former focusing on the discovery from a reference attribute, while the latter fully searching all attributes in a reference dataset.

\begin{definition}[Discovery-by-attribute]
Let $A_q$ be a query attribute, $D_{ref}$ a reference dataset where $A_q \in att(D_{ref})$, and $\mathcal{D}$ a dataset repository where $D_{ref} \notin \mathcal{D}$; obtain a partially-ordered set of equi-join predicates $\mathcal{P}$ of the form $\{ \langle A_q, A_1 \rangle, $ \ldots, $ \langle A_q, A_n \rangle \}$, where $A_1, $ \ldots, $ A_n \in att(\mathcal{D})$ such that $\forall \langle A_q, A_i \rangle, \langle A_q, A_j \rangle \in \mathcal{P}: \langle A_q, A_i \rangle \succ \langle A_q, A_j \rangle \implies Predict(P_u(A_q), P_u(A_i), P_b(A_q,A_i))$ $\geq Predict($ $P_u(A_q), P_u(A_j),$ $ P_b(A_q,A_j))$.
\end{definition}

\begin{definition}[Discovery-by-dataset]
Let $D_{ref}$ be a reference dataset, and $\mathcal{D}$ a dataset repository where $D_{ref} \notin \mathcal{D}$; obtain a partially-ordered set of equi-join predicates $\mathcal{P}$ of the form $\{ \langle A^\prime, A_1 \rangle,$ $ \ldots$, $\langle A^{\prime\prime}, A_n \rangle \}$,  where $A^\prime, A^{\prime\prime} \in att(D_{ref})$ and $A_1, \ldots,$ $A_n$ $\in att(\mathcal{D})$, such that $\forall \langle A^\prime, A_i \rangle, \langle A^{\prime\prime}, A_j \rangle \in \mathcal{P}: \langle A^\prime, A_i \rangle \succ \langle A^{\prime\prime}, A_j \rangle$ $\implies Predict(P_u(A^\prime),$ $ P_u(A_i), P_b(A^\prime,A_i)) \geq Predict($ $P_u(A^{\prime\prime}),$ $P_u(A_j)$ ,  $P_b(A^{\prime\prime},A_j))$.
\end{definition}

From each proposed and selected equi-join predicate in each scenario, we automatically generate a conjunctive query (see Section \ref{sec:datamodel}) to enable an informed data analysis.

\section{A learning approach to join discovery}\label{sec:approach}

In this section, we describe in detail our approach. First, we present how we model and construct profiles, in order to later build learning models that allow to accurately predict joins.

\subsection{Attribute profiling}

Profiles are composed of meta-features that represent the underlying characteristics of attributes. Such profiles are the key ingredient for high accuracy predictions, thus we require an exhaustive summary of attributes. To this end, we base our profiling on state-of-the-art relational data profiling techniques \cite{DBLP:journals/vldb/AbedjanGN15}. We distinguish meta-features corresponding to unary and binary profiles. We further distinguish the former into meta-features modeling cardinalities, value distribution and syntax. A summary of all the implemented meta-features is depicted in Table \ref{tab:metadataList}. Note that, although for space reasons it has not been included here, we validated through a principal component analysis the relevance of all meta-features towards meaningful profiling of attributes.

\begin{table*}[htbp!]
	\centering
	\resizebox{\textwidth}{!}{
		\begin{tabular}{|c|c|l|c|}
			\hline
			\textbf{Category} & \textbf{Meta-feature} & \multicolumn{1}{c|}{\textbf{Description}} & \textbf{Norm.?} \\
			\hline
			\multirow{4}{*}{Cardinalities}                                                  & Cardinality              & Number of distinct values within an attribute & Yes                                                                                                \\
			\cline{2-4}
			& Uniqueness & Measures if the attribute contains unique values   & No \\
			\cline{2-4}
			& Incompleteness           & Measures the number of missing values                                                                                                      & No                                                                                                 \\
			\cline{2-4}
			& Entropy                  & Measures the variety of an attribute                                                                                                  & Yes                                                                                                \\
			\hline
			\multirow{11}{*}{\begin{tabular}[c]{@{}c@{}}Value \\distribution \end{tabular}} & Average frequency        & The average value of the frequency distribution count                                                                                      & Yes                                                                                                \\
			\cline{2-4}
			& Min frequency            & The minimum value of the frequency distribution count                                                                                      & Yes                                                                                                \\
			\cline{2-4}
			& Max frequency            & The maximum value of the frequency distribution count                                                                                      & Yes                                                                                                \\
			\cline{2-4}
			& SD frequency             & The standard deviation of the frequency distribution count                                                                                 & Yes                                                                                                \\
			\cline{2-4}
			& Octiles                  & The octiles (quantiles) of the frequency distribution in percentages                                                                       & No                                                                                                 \\
			\cline{2-4}
			& Min perc frequency       & The minimum value of the frequency distribution in percentages                                                                             & No                                                                                                 \\
			\cline{2-4}
			& Max perc frequency       & The maximum value of the frequency distribution in percentages                                                                             & No                                                                                                 \\
			\cline{2-4}
			& SD perc frequency        & The standard deviation of the frequency distribution in percentages                                                                        & No                                                                                                 \\
			\cline{2-4}
			& Constancy                & Frequency of the most frequent value divided by number of rows                                                                             & No                                                                                                 \\
			\cline{2-4}
			& Frequent words           & The 10 most frequent words                                                                                                                 & No                                                                                                 \\
			\cline{2-4}
			& Soundex                  & The 10 most frequent words in soundex representation                                                                                       & No                                                                                                 \\
			\hline
			\multirow{12}{*}{Syntactic}                                                      & Data type                & \begin{tabular}[c]{@{}l@{}}The data type of the attribute (i.e., numeric, alphanumeric, alphabetic, \\nonAlphanumeric, or datetime) \end{tabular}    & No                                                                                                 \\
			\cline{2-4}
			& Specific type            & \begin{tabular}[c]{@{}l@{}}The specific type of the attribute (i.e., phone, email, url, ip, username, or phrases) \end{tabular}                    & No                                                                                                 \\
			\cline{2-4}
			& Percentage data type     & The percentage for each data type detected in the data values                                                                              & No                                                                                                 \\
			\cline{2-4}
			& Percentage specific type & The percentage for each specific type detected in the data values                                                                          & No                                                                                                 \\
			\cline{2-4}
			& Longest string           & The number of characters in the longest string                                                                                             & Yes                                                                                                \\
			\cline{2-4}
			& Shortest string          & The number of characters in the shortest value in the attribute                                                                            & Yes                                                                                                \\
			\cline{2-4}
			& Average string           & Average length of the strings in term of characters                                                                                        & Yes                                                                                                \\
			\cline{2-4}
			& Number words             & The number of words in the attribute                                                                                                       & Yes                                                                                                \\
			\cline{2-4}
			& Average words            & The average words in the attribute                                                                                                         & Yes                                                                                                \\
			\cline{2-4}
			& Min words                & The minimum words in the attribute                                                                                                         & Yes                                                                                                \\
			\cline{2-4}
			& Max words                & The maximum words in the attribute                                                                                                         & Yes                                                                                                \\
			\cline{2-4}
			& SD words                 & The standard deviation in the attribute                                                                                                    & Yes                                                                                                \\
			\hline
			\multirow{3}{*}{\begin{tabular}[c]{@{}c@{}}Pair\\metadata \end{tabular}}      & Best containment          & The containment score assuming all distinct values are covered                                                                           & No                                                                                                 \\
			\cline{2-4}
			& Flipped containment       & \begin{tabular}[c]{@{}l@{}}Containment assuming all distinct values are covered divided by max cardinality \end{tabular} & No                                                                                                 \\
			\cline{2-4}
			& Name distance  & Measures the difference of two attribute names using Levenshtein distance & No                                                                                                 \\
			\hline
		\end{tabular}
	}
	\captionof{table}{Meta-features composing a profile} \label{tab:metadataList}
\end{table*}

\medskip

\noindent\textbf{Cardinalities.} These meta-features provide a broad view of an attribute. Uniqueness, which is computed dividing the number of distinct values by the cardinality, allows us to quantify the extent of duplicated values. A uniqueness smaller than $1$ indicates there exists duplicate values, hence we can identify which attributes have high redundancies. We can also detect incompleteness, which is determined by the number of missing values divided by the cardinality. This produces a value in the range $[0,1]$, where values closer to $1$ denote the attribute has a high percentage of missing values. Finally, entropy, also referred as \textit{diversity index}, measures the variety of data in an attribute. 

\medskip

\noindent\textbf{Value distribution.}
Here, we exploit information in a fine-grained manner by using a frequency distribution of the attribute values, either by count or percentage. Despite its simplicity, the frequency distribution of column values exposes insightful characteristics, such as how often a value occurs. We compute frequency metrics (e.g., in the form of octiles), and descriptive statistics (e.g., mean, standard deviation, etc.) to characterize the distribution of the data. We also take a sample of the ten most frequent values.

\medskip

\noindent\textbf{Syntax.}
This category of unary metadata describes the shape of data and their pattern. We consider that looking for patterns in data provides an abstract representation of the attribute. These meta-features include information regarding the length of values in characters, such as the length of the longest and shortest value, and the average length. We also compute information regarding syntactic consistency, such as format and data type. This aids to give meaning to the attribute's content. We also infer the data type of an attribute, in a broad and fine-grained manner. Broad data types are generic descriptions such as numeric, alphabetic, alphanumeric, dateTime, non-alphanumeric, etc. However, we also extract its fine-grained type to extract what content is the attribute representing. To this end, we use regular expressions that allow us to model usernames (i.e., strings with length smaller than six characters), phrases (i.e., strings larger than six characters), phones, emails, URLs, IPs, etc. In order to improve the quality of meta-features in this category, we preprocess values to lowercase, remove accents and special symbols. 

\medskip

\noindent\textbf{Binary meta-features.}
We also extract meta-features regarding pairs of attributes. We use Levenshtein distance to obtain the similarity between pairs of attribute names \cite{1966SPhD...10..707L}. This is normalized by the length of the largest string. We can also extract an estimation of the ratio for a possible join, that is, the number of tuples that two attributes can generate, depicted \textit{best containment} ($C_B$). Since we do not perform a comparison-by-value, we obtain this estimation using the cardinality from each attribute, assuming the best scenario where all distinct values are covered in both attributes. This will allow us to identify which pair of attributes can produce the most desirable quality join since not all pair of attributes will produce the same ratio. We also compute its opposite metric, which we denote \textit{flipped containment} ($C_F$). This is an indicator of the cardinality proportion w.r.t. the attribute with the smallest cardinality. These are formalized as follows:

\medskip

\begin{minipage}{.49\linewidth}
	\centering
	\begin{equation*} \label{eq:bestContainment}
	C_B(A,B) = \frac{ min(|A|,|B|)}{|A|}
	\end{equation*}
\end{minipage}
\begin{minipage}{0em}
	\begin{equation*} \label{eq:flippedContainment}
	C_F(A,B) = \frac{ min(|A|,|B|)}{max(|A|,|B|)}
	\end{equation*}
\end{minipage}

\subsection{Comparing profiles}

Before comparing profiles and due to the fact attribute meta-features are represented in different magnitudes, we normalize them to guarantee a meaningful comparison. As shown in Table \ref{tab:metadataList}, we consider a large amount of meta-features that require normalization. Two common normalization techniques are Min-Max and Z-score. The former consists on rescaling data into the range $[0,1]$, This technique, however, is sensitive to outliers which will lay on the boundaries. Oppositely, Z-score normalization overcomes this issue by rescaling values to have a mean of $0$ and a standard deviation of $1$. For this reason, we use Z-score to normalize meta-features. The following equation depicts the normalization process, which requires the mean and standard deviation of the metadata, which is computed from all the values of each attribute to be compared.

\mathchardef\mhyphen="2D
\begin{equation*} \label{eq:z_score}
Z\mhyphen score = \frac{(x - \mu )}{\sigma }\
\end{equation*}


After normalizing each meta-feature we compute the distances among pairs of attributes. Here, we also compute binary meta-features. The result of this stage is a set of distance vectors $D$ where, for each $D_i$, values closer to $0$ denote high similarities.

\subsection{Predictive model}\label{sec:predictiveModel}

Once the distance vectors are computed, we can train the predictive model. Precisely, the goal is to train a model so that, for a pair of attributes $A,B$, its prediction (i.e., $Predict(A,B)$) is highly correlated to the true class (i.e., $Quality(A,B)$). To this end, we deploy a random forest classifier. This is a highly expressive model, which is also robust to outliers. Here, we present the methodology we used to generate it.

\medskip

\begin{figure}[h!]
	\begin{center}
		\includegraphics[width=1\linewidth]{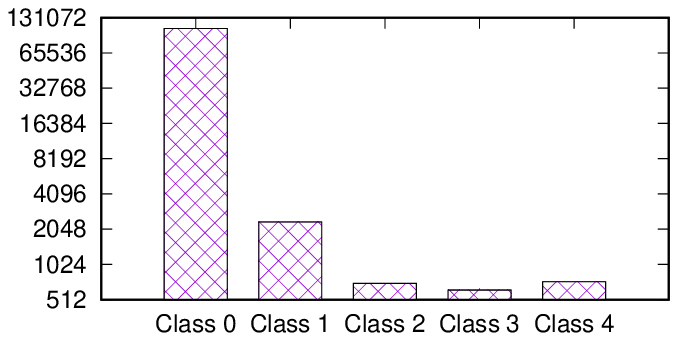}
		\vspace{-2em}
		\caption{Frequency distribution for each class in the training set using a base 2 log scale}
		\label{fig:distributionTraining}
	\end{center}
\end{figure}

\noindent\textbf{Ground truth.}
We selected 138 datasets from open repositories such as Kaggle\footnote{\url{https://www.kaggle.com/}} and OpenML\footnote{\url{https://www.openml.org/}} sites\footnote{The complete ground truth repository is available at \url{https://mydisk.cs.upc.edu/s/GeYwdYH7xsGqbaX}}. Precisely, we devised an heterogeneous collection of datasets ranging different topics that could generate high quality joins. Such collection allowed us to generate a training data set of $110,378$ candidates pairs of string attributes, where $4404$ of those are joinable attributes with qualities ranging from $1$ to $4$. Recall from the definition of $Quality$, that higher quality labels entail higher restrictions on the containment and cardinality proportion. This, creates an imbalance problem, as shown in Figure \ref{fig:distributionTraining}, where the majority of the sample falls into class $0$, leaving underrepresented the rest of classes.

\medskip

\begin{figure}[h!]
	\begin{center}
		\includegraphics[width=1\linewidth]{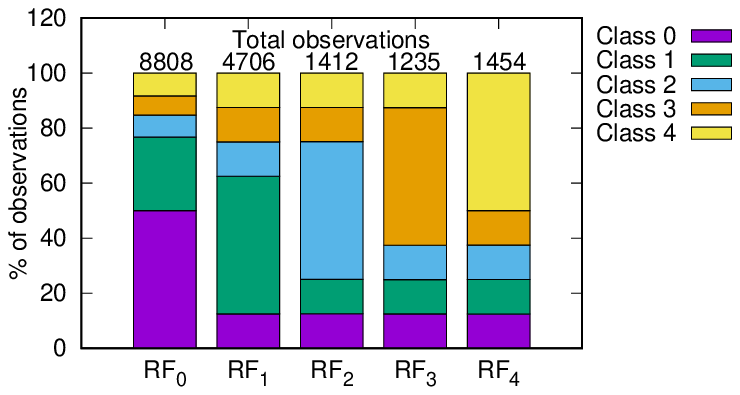}
		\vspace{-2em}
		\caption{One-vs.-rest frequency distribution per class}
		\label{fig:ovrDistribution}
	\end{center}
\end{figure}

\noindent\textbf{One-vs.-rest.}
In order to generate a classifier not affected by imbalanced classes we adopt a \textit{one-vs.-rest} strategy. This is a method that splits a multi-class classification problem into a binary classification problem per class, and maximizing the number of training data per class. Hence, the classifier predicting the quality class $S_0$, is fed with $S_0$ positive pairs and the same amount of $S_1, \ldots, S_4$ negative pairs (preserving, when possible, a uniform distribution of each negative class). The process is likewise for the other classes. The most notorious impact of this strategy is that each classifier is trained with an optimal number of elements for the positive class. The negative class is chosen so that it maximizes the representativeness of the other classes. Figure \ref{fig:ovrDistribution}, depicts the balanced frequency distribution of this strategy. For each $RF_i$, observations for class $i$ are considered positive, while all the others are negative.

\medskip

\noindent\textbf{Chain classifier.}
In order to reduce the false positive rate,the different classifiers are connected in a classifier chain architecture. This is a well-studied and effective approach for multi-label classification \cite{DBLP:journals/ml/ReadPHF11}. Each classifier predicting the probability of class $i$ is trained with the set of distance vectors $D$ and the probabilities of classes $0, \ldots, i-1$. This approach improves the predictive accuracy of the classifier, as we show in Section \ref{sec:accuracy}. Figure \ref{fig:chainclassifier} depicts a high-level overview of the architecture used for training. Then, the prediction returned by the classifier is the one with highest probability from each $RF_i$.

\medskip

\medskip

\noindent\begin{minipage}{\linewidth}
	\centering
    \begin{tikzpicture}[
        circle/.style = {shape=circle,rounded corners,draw,minimum size=2em,font=\Large},
        rect/.style = {shape=rectangle,rounded corners,draw,minimum size=2em},
        edge/.style = {- = latex'},
        to/.style={->,>=stealth',shorten >=1pt}]
		
		\node[rect] (RECT0) at (0,2.5) {$D$};
		\node[circle,accepting] (RF0) at (0,1.25) {$RF_0$};
		\node[rect] (RECT1) at (1.6,0) {$D, p_0$};
		\node[circle,accepting] (RF1) at (1.6,1.25) {$RF_1$};
		\node[rect] (RECT2) at (3.3,2.5) {$D, p_0, p_1$};
		\node[circle,accepting] (RF2) at (3.3,1.25) {$RF_2$};
		\node[rect] (RECT3) at (5,0) {$D, p_0, p_1, p_2$};
		\node[circle,accepting] (RF3) at (5,1.25) {$RF_3$};
		\node[rect] (RECT4) at (6.7,2.5) {$D, p_0, p_1, p_2, p_3$};
		\node[circle,accepting] (RF4) at (6.7,1.25) {$RF_4$};
		
		\draw[to] (RECT0) to[out=south,in=north] (RF0); 
		
		\draw[to] (RF0) to[out=south,in=west] (RECT1); 
		\draw[to] (RECT1) to[out=north,in=south] (RF1);
		
		\draw[to] (RF1) to[out=north,in=west] (RECT2); 
		\draw[to] (RECT2) to[out=south,in=north] (RF2); 
		
		\draw[to] (RF2) to[out=south,in=west] (RECT3); 
		\draw[to] (RECT3) to[out=north,in=south] (RF3); 
		
		\draw[to] (RF3) to[out=north,in=west] (RECT4); 
		\draw[to] (RECT4) to[out=south,in=north] (RF4); 
	\end{tikzpicture}
	
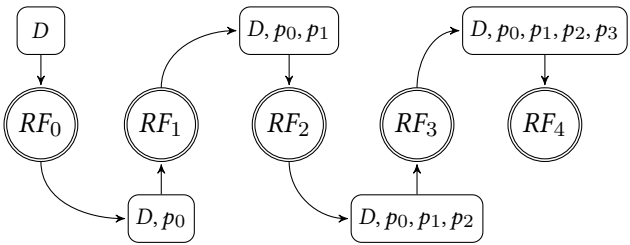
\captionof{figure}{Chain of 5 random forest classifiers, each $RF_i$ is fed with the distances vectors $D$ and the probabilities from the previous classifiers $p_0, \ldots, p_{i-1}$}
	\label{fig:chainclassifier}
\end{minipage}

\medskip

\noindent\textbf{Hyperparameter tuning.}
We perform a model-specific hyperparameter search \cite{DBLP:journals/widm/ProbstWB19}. For each $RF_i$, we perform a $k$-fold cross-validation training ($k=5$), using an $80\%/20\%$ training/set partition. We, precisely, tune the \textit{maxBins}, \textit{maxDepth} and the number of trees. We do not tune other hyperparameters (e.g., feature subset strategy or impurity). We tried the following sets of values: $maxBins = \{35,40,45,50,55,60\}$, $maxDepth = \{ 20, 25, 30\}$, and $trees = \{ 64,$ $70, 80, 90, 100, 110, 120,$ $128 \}$. Table \ref{tab:hyperparameters}, shows the resulting hyperparameter choice and the training time, which was executed on an \textit{n1-standard-8} machine hosted by the Google Cloud Platform.

\begin{table}[h!]\small
	\centering
	\resizebox{0.45\textwidth}{!}{
		\begin{tabular}{|c|c|c|c|c|c|}
			\hline
			& $\bm{RF_0}$ & $\bm{RF_1}$ & $\bm{RF_2}$ & $\bm{RF_3}$ & $\bm{RF_4}$  \\
			\hline
			MaxBins & 55 & 40 & 40 & 60 & 45 \\
			\hline
			MaxDepth & 25 & 25 & 30 & 30 & 20  \\
			\hline
			Number of trees & 120 & 100 & 120 & 80 & 120 \\
			\hline
			Training time (hrs) & 2.92 & 4.03 & 4.42 & 4.98 & 3.89 \\
			\hline
		\end{tabular}
	}
	\caption{\label{tab:hyperparameters} Selected hyperparameters per model}
\end{table}

\noindent\textbf{Ranking.}
Finally, in order to assign the predicted quality class to a candidate attribute, we initially assign the label with the highest probability from all classifiers. However, some times, the highest probability is close to some of the other class probabilities. Thus, we complement the prediction taking into account the class 0 probability $p_0$ (i.e., the probability candidate pair is not joinable) and considering the following two scenarios, where $p_0$ will affect the predicted class we finally assign:

\begin{enumerate}[leftmargin=1.5em]
\item \label{dataFusion.case1} Assume that $p_0 > 0.5$, however the maximum probability belongs to any $p_1, \ldots, p_4$. Since $p_0$ denotes no join predicted, we downgrade the predicted class from $i$ to the class in $0, \ldots, i-1$ with the second highest probability.
\item \label{dataFusion.case2} The second scenario occurs when all $p_0, \ldots, p_4$ have a probability smaller than $0.5$, where we consider that the candidate pair is prone to misclassification. In this situation, if the higher probability belongs to any $p_1, \ldots, p_4$, but it still close to $p_0$ (i.e., it does not exceed a threshold\footnote{In our experiments we empirically defined the threshold to be $0.10$.}), then we downgrade the predicted class from $i$ to the class in $0, \ldots, i-1$ with the second highest probability.
\end{enumerate}

\section{Evaluation}\label{sec:evaluation}

In this section, we present the evaluation of our approach. On the one hand, we evaluate the ability of the model to discover quality joins through several experiments as well as its generalizability. On the other hand, we compare its performance with representative state-of-the-art solutions. In order to present transparent experiments and guarantee the reproducibility of results, we created an informative companion website\footnote{{\url{https://www.essi.upc.edu/dtim/nextiajd/}}}. There, it is possible to find all necessary resources (i.e., source code, datasets, and detailed instructions) needed to reproduce the presented results. The companion website also provides access to \name, a tool implementing our approach, as well as a ready-to-use demo in the form of a Jupyter notebook running \name.

\subsection{Setting}

\noindent\textbf{Implementation.}
We have implemented \name~as an extension of Apache Spark 3.0.1. The classification model was trained using its distributed machine learning library MLlib. The runtime methods (i.e., profiling, and generation of rankings ordered by the predicted join quality) are implemented as new operators over the structured data processing library SparkSQL. We leverage on the Catalyst optimizer to efficiently compute the profiles and compare them. As described in the problem statement in Section \ref{sec:statement}, our implementation supports two modes of operation \textit{discovery-by-attribute} and \textit{discovery-by-dataset}. The former receives as input a reference Spark dataframe (i.e., a dataset) and one of its attributes, and generates a ranking against a collection of dataframes. The \textit{discovery-by-dataset} mode does not receive as input a reference attribute, so it runs the discovery process for all attributes from the reference dataframe.
Notably, implementing \name~on top of Spark brings many other benefits. Firstly, we can benefit from many source connectors and we can easily ingest the most common data formats (e.g., CSV, JSON, XML, Parquet, Avro, etc.). Secondly, our extension benefits from the inherent capacity of Spark to parallelize tasks on top of distributed data. 

\medskip

\noindent\textbf{Test set.}
For evaluation purposes, we collected 139 independent datasets from those used for the ground truth. We further divided such datasets into 4 testbeds (extra-small, small, medium and large) according to their file size. Table \ref{tab:testbeds} shows the characteristics of each testbed.

\begin{table}[h!]
	\centering
	\resizebox{0.475\textwidth}{!}{
		\begin{tabular}{|c|c|c|c|c|}
			\hline
			\textbf{Testbed} & $\bm{XS}$ & $\bm{S}$ & $\bm{M}$ & $\bm{L}$  \\
			\hline
			File size & $0-1$ MB & $1 - 100$ MB & $100$ MB $- 1$ GB & $>1$ GB \\
			\hline
			Datasets & 28 & 46 & 46 & 19 \\
			\hline
			\begin{tabular}[c]{@{}c@{}}String\\attributes \end{tabular} & 159 & 590 & 600 & 331 \\
			\hline
		\end{tabular}
	}
	\caption{\label{tab:testbeds} Characteristics per testbed}
\end{table}


\medskip

\noindent\textbf{Alternatives.}
We compare our approach with the following state-of-the-art data discovery solutions, whose source code is openly available: LSH Ensemble \cite{DBLP:journals/pvldb/ZhuNPM16} and FlexMatcher \cite{DBLP:journals/debu/ChenGHTD18}. No fine tuning was performed in such systems, running the code as provided out-of-the-box. These systems differ in their mode of operation, as described in Section \ref{sec:relatedwork}, the former being an approach based on comparison by hash, while the latter on comparison by profile. Note we rule out approaches based on comparison by value. Such solutions are not comparable to ours, since their kind of search accuracy is exact and by nature they are not suitable for large-scale scenarios.

\medskip

\noindent\textbf{Setup.}
We distinguish two experimental setups. Given that the other systems are not implemented to be executed in a distributed and parallel fashion. In order to perform a fair comparison, the experiments focusing on the quality of the results obtained are run on a quad-core 2018 Macbook Pro 2.30GHz, with a 256GB SSD disk and 16 GB of RAM memory.
Then, the experiments that stress the scalability of our method are run on an InfiniBand network 4-node cluster (1 driver, 3 workers), where each machine is equipped with a 1TB SSD disk, 16 CPU 2.27GHz cores and 48GB of RAM memory.

\setcounter{figure}{4}
\begin{figure*}[!b]
	\begin{center}
		\begin{subtable}{.225\linewidth}
			\centering
			\caption{\label{tab:matrixXS}\hspace{1em} $XS$ ($0-1$ MB)}
			\[
			\raisebox{-2\baselineskip}{%
				\begin{tabular}[b]{c@{\hspace*{0.5em}}c} \vspace{-3em}
				& {}
				\\
				\begin{sideways}
				\phantom{0}True class
				\end{sideways}
				& \raisebox{0.5\height}%
				{\(
					\begin{matrix}
					{} & \phantom{00} & \phantom{00} & \phantom{00} & \phantom{00} & \phantom{00} \\
					\textbf{4} & \cellcolor{d5}0 & \cellcolor{d4}0 & \cellcolor{d3}6 & \cellcolor{d2}27 & \cellcolor{d1}52 \\
					\textbf{3} & \cellcolor{d4}0 & \cellcolor{d3}0 & \cellcolor{d2}3 & \cellcolor{d1}6 & \cellcolor{d2}0 \\
					\textbf{2} & \cellcolor{d3}2 & \cellcolor{d2}0 & \cellcolor{d1}45 & \cellcolor{d2}4 & \cellcolor{d3}0 \\
					\textbf{1} & \cellcolor{d2}16 & \cellcolor{d1}8 & \cellcolor{d2}7 & \cellcolor{d3}1 & \cellcolor{d4}0 \\
					\textbf{0} & \cellcolor{d1}\app & \cellcolor{d2}44 & \cellcolor{d3}68 & \cellcolor{d4}0 & \cellcolor{d5}1 \\
					{} & \textbf{0} & \textbf{1} & \textbf{2} & \textbf{3} & \textbf{4}
					\end{matrix}
					\)%
				} \\
				& Predicted class
				\end{tabular}}
			\]
		\end{subtable}
		\begin{subtable}{.01\linewidth}
			\hspace*{\fill}
		\end{subtable}
		\begin{subtable}{.245\linewidth}
			\caption{\label{tab:matrixS}\hspace{1em} $S$ ($1 - 100$ MB)}
			\centering
			\[ 
			\raisebox{-2\baselineskip}{%
				\begin{tabular}[b]{c@{\hspace*{0.5em}}c} \vspace{-3em}
				& {}
				\\
				& \raisebox{0.5\height}%
				{\(
					\begin{matrix}
					& \phantom{00} & \phantom{00} & \phantom{00} & \phantom{00} & \phantom{00} \\
					\cellcolor{d5}2 & \cellcolor{d4}0 & \cellcolor{d3}19 & \cellcolor{d2}116 & \cellcolor{d1}161 \\
					\cellcolor{d4}1 & \cellcolor{d3}3 & \cellcolor{d2}67 & \cellcolor{d1}149 & \cellcolor{d2}1 \\
					\cellcolor{d3}44 & \cellcolor{d2}1 & \cellcolor{d1}93 & \cellcolor{d2}10 & \cellcolor{d3}0 \\
					\cellcolor{d2}1553 & \cellcolor{d1}390 & \cellcolor{d2}197 & \cellcolor{d3}0 & \cellcolor{d4}0 \\
					\cellcolor{d1}\app & \cellcolor{d2}1718 & \cellcolor{d3}394 & \cellcolor{d4}6 & \cellcolor{d5}0 \\
					\textbf{0} & \textbf{1} & \textbf{2} & \textbf{3} & \textbf{4}
					\end{matrix}
					\)%
				} \\
				& Predicted class
				\end{tabular}}
			\]
		\end{subtable}%
		\begin{subtable}{.01\linewidth}
			\hspace*{\fill}
		\end{subtable}
		\begin{subtable}{.235\linewidth}
			\caption{\label{tab:matrixM}\hspace{1em} $M$ ($100$ MB $- 1$ GB)}
			\centering
			\[ 
			\raisebox{-2\baselineskip}{%
				\begin{tabular}[b]{c@{\hspace*{0.5em}}c} \vspace{-3em}
				& {}
				\\
				& \raisebox{0.5\height}%
				{\(
					\begin{matrix}
					\phantom{00} & \phantom{00} & \phantom{00} & \phantom{00} & \phantom{00} \\
					\cellcolor{d5}1 & \cellcolor{d4}0 & \cellcolor{d3}16 & \cellcolor{d2}110 & \cellcolor{d1}134 \\
					\cellcolor{d4}4 & \cellcolor{d3}3 & \cellcolor{d2}41 & \cellcolor{d1}169 & \cellcolor{d2}0 \\
					\cellcolor{d3}46 & \cellcolor{d2}2 & \cellcolor{d1}96 & \cellcolor{d2}56 & \cellcolor{d3}1 \\
					\cellcolor{d2}3017 & \cellcolor{d1}242 & \cellcolor{d2}53 & \cellcolor{d3}0 & \cellcolor{d4}0 \\
					\cellcolor{d1}\app & \cellcolor{d2}508 & \cellcolor{d3}142 & \cellcolor{d4}50 & \cellcolor{d5}0 \\
					\textbf{0} & \textbf{1} & \textbf{2} & \textbf{3} & \textbf{4}
					\end{matrix}
					\)%
				} \\
				& Predicted class
				\end{tabular}}
			\]
		\end{subtable}%
		\begin{subtable}{.01\linewidth}
			\hspace*{\fill}
		\end{subtable}
		\begin{subtable}{.23\linewidth}
			\caption{\label{tab:matrixL}\hspace{1em} $L$ ($>1$ GB)}
			\centering
			\[ 
			\raisebox{-2\baselineskip}{%
				\begin{tabular}[b]{c@{\hspace*{0.5em}}c} \vspace{-3em}
				& {}
				\\
				& \raisebox{0.5\height}%
				{\(
					\begin{matrix}
					\phantom{00} & \phantom{00} & \phantom{00} & \phantom{00} & \phantom{00} \\
					\cellcolor{d5}2 & \cellcolor{d4}2 & \cellcolor{d3}27 & \cellcolor{d2}46 & \cellcolor{d1}56 \\
					\cellcolor{d4}2 & \cellcolor{d3}1 & \cellcolor{d2}40 & \cellcolor{d1}24 & \cellcolor{d2}3 \\
					\cellcolor{d3}0 & \cellcolor{d2}14 & \cellcolor{d1}78 & \cellcolor{d2}28 & \cellcolor{d3}0 \\
					\cellcolor{d2}2317 & \cellcolor{d1}97 & \cellcolor{d2}92 & \cellcolor{d3}0 & \cellcolor{d4}0 \\
					\cellcolor{d1}\app & \cellcolor{d2}100 & \cellcolor{d3}56 & \cellcolor{d4}0 & \cellcolor{d5}0 \\
					\textbf{0} & \textbf{1} & \textbf{2} & \textbf{3} & \textbf{4}
					\end{matrix}
					\)%
				} \\
				& Predicted class
				\end{tabular}}
			\]
		\end{subtable}%
	\end{center}
	\vspace{-1em}
	\caption{ \label{fig:confusionMatricesPerCategory} Confusion matrices per file size category. For spacing reasons, values in the bottom left corner are omitted}
\end{figure*}

\subsection{Predictive performance}\label{sec:accuracy}

The goal of the first experiment is to assess the classifier's predictive performance by generating and evaluating the ranking of candidate equi-join predicates for each testbed. We also aim at studying whether the file size impacts the quality of the prediction. With these experiments, we aim at proving the generalizability of our proposal. 

\medskip

\noindent\textbf{Methodology.}
We will use a confusion matrix to capture the relationship between the true and predicted classes. Additionally, we provide performance metrics for the classifier such as precision, recall and $F_1$ score. We first discuss the experiment for all testbeds together, and later do a fine-grained discussion for each testbed.

\medskip

\noindent\textbf{Results.}
Figure \ref{fig:confusionMatrixAll} and Table \ref{tab:accuracyrecall}, show, respectively, the confusion matrix and performance metrics for all testbeds. Overall, we evaluated 467965 attributes pairs. We can validate the good performance of the proposed approach by the fact that class 4, denoting the highest quality joins, has the best precision. Our method aims at proposing a ranking according to the predicted join quality, which for the highest value it has almost no false positives. It is also relevant to note that the prediction for class 0, denoting the no join quality, also have both high precision and recall. This is particularly relevant to filter out uninteresting results, and thus reduce the search space when presenting a ranking to the user. We additionally note that, as depicted by the precision and recall measures, predictions corresponding to classes 1 and 2 are highly inaccurate.
\setcounter{figure}{3}
\begin{figure}[!h]
	\vspace{-2em}
	\[ 
	\raisebox{-2\baselineskip}{%
		\begin{tabular}[b]{c@{\hspace*{0.5em}}c}
		& {}
		\\
		\begin{sideways}
		\phantom{0}True class
		\end{sideways}
		& \raisebox{0.5\height}%
		{\(
			\begin{matrix}
			{} & \phantom{000000} & \phantom{000000} & \phantom{000000} & \phantom{000000} & \phantom{000000} \\
			\textbf{4} & \cellcolor{d5}5 & \cellcolor{d4}2 & \cellcolor{d3}68 & \cellcolor{d2}299 & \cellcolor{d1}393 \\
			\textbf{3} & \cellcolor{d4}7 & \cellcolor{d3}7 & \cellcolor{d2}151 & \cellcolor{d1}348 & \cellcolor{d2}4 \\
			\textbf{2} & \cellcolor{d3}92 & \cellcolor{d2}17 & \cellcolor{d1}312 & \cellcolor{d2}98 & \cellcolor{d3}1 \\
			\textbf{1} & \cellcolor{d2}6903 & \cellcolor{d1}737 & \cellcolor{d2}349 & \cellcolor{d3}1 & \cellcolor{d4}0 \\
			\textbf{0} & \cellcolor{d1}455084 & \cellcolor{d2}2370 & \cellcolor{d3}660 & \cellcolor{d4}56 & \cellcolor{d5}1 \\
			{} & \textbf{0} & \textbf{1} & \textbf{2} & \textbf{3} & \textbf{4}
			\end{matrix}
			\)%
		} \\
		& Predicted class
		\end{tabular}}
	\]
	\vspace{-1.5em}\captionof{figure}{\label{fig:confusionMatrixAll}Confusion matrix for all testbeds (clearer cells denote a closer proximity w.r.t. the true class)}
\end{figure}
\setcounter{figure}{5}
\begin{table}[!h]
	\centering
	\resizebox{0.4\textwidth}{!}{
		\begin{tabular}{|c|c|c|c|}
			\hline
			 & \textbf{Precision} & \textbf{Recall} & \textbf{$\bm{F_1}$ score} \\
			\hline
			$(0)\text{ }\texttt{None}$ & $0.9848$ & $0.9933$ & $0.9890$ \\
			\hline
			$(1)\text{ }\texttt{Poor}$ & $0.2352$ & $0.0922$ & $0.1325$ \\
			\hline
			$(2)\text{ }\texttt{Moderate}$ & $0.2025$ & $0.6000$ & $0.3029$ \\
			\hline
			$(3)\text{ }\texttt{Good}$ & $0.4339$ & $0.6731$ & $0.5276$ \\
			\hline
			$(4)\text{ }\texttt{High}$ & $0.9849$ & $0.5123$ & $0.6740$ \\
			\hline
		\end{tabular}
	}
	\caption{\label{tab:accuracyrecall} Performance metrics per class for all testbeds}
\end{table}
Nevertheless, \name~is prepared to be used as an interactive tool. Thus, if we analyze these results from the point of view of a user, most misclassifications are irrelevant. \name~shows the results in strict order. First, classes 4 and 3, and then classes 2 and 1 on demand. In this sense, a binary classification meaning \emph{likely interesting} or \emph{likely uninteresting} would be a fairer way to evaluate~\name. When considering these results as a binary problem (interesting: classes 3-4; uninteresting: classes 0-2), the evaluated metrics improve considerably, as shown in Figure \ref{fig:confusionMatricesBinaryNexiavsLSH} and Table \ref{tab:allSystemsAccuracyRecall}. Relevantly, we note that \name~generates very few false positives; a desirable property for large-scale data discovery problems. We nevertheless highlight the relevance of distinguishing classes 1 and 2 from 0, either for advanced users or because \name~could be used for other automatic data discovery problems where such distinction would be relevant. Finally, we analyzed the semantic validity of the join pairs proposed as interesting. Relevantly, 95\% of the attribute pairs shown to the user have a semantic relationship.

Next, in Figure \ref{fig:confusionMatricesPerCategory}, we present a fine-grained evaluation of the predictions per file category. Overall, we confirm 
that regardless of the file size, we accurately predict quality joins. In all cases, we miss between 40\% and 50\% of class 4 predictions, due to the fact they are downgraded to lower quality classes (mostly to 3). This is the reason for the low recall for that class. However, this is not a concerning aspect, as most candidates will still appear in the ranking but in lower positions. We again highlight the low ratio of false positives generated. 
\subsection{Comparison with the state-of-the-art}\label{sec:comparisonSotA}

In this experiment we aim at comparing our approach to other data discovery approaches. We perform such evaluation by measuring and comparing their computational complexity and predictive performance. 

\medskip

\noindent\textbf{Methodology.}
All systems under evaluation, including ours, implement data discovery in two steps. The first step, which we denote \textit{pre}, builds the core data structures from the datasets. For hash-based methods, such as LSH Ensemble, building the index, while profile-based methods, such as FlexMatcher and ours, create the profiles. Additionally, FlexMatcher will create the predictive models for each new data discovery task. Then, the second step, which we denote as \textit{query}, consists of computing the prediction leveraging on the previously built data structures. Whenever possible, we decouple both steps and thus read the data structures from disk. This is, however, not the case for LSH Ensemble, as it does not offer any resources to store the index on disk, forcing us to maintain it in memory.

The three systems analyzed have slightly different objectives. In order to perform a fair comparison, we analyze the results from the user perspective. That is, the number of results provided and its degree of interestingness for the user. For that, we use a binary scale, which straightforwardly maps to the output obtained in LSH Ensemble and FlexMatcher. For \name, we will reuse the interestingness mapping discussed in the previous experiment: we map classes $\{ 0,1,2 \}$ to the uninteresting class, and classes $\{ 3,4 \}$ to interesting. Applying the same rationale, in LSH Ensemble we consider interesting those pairs with a containment above $50\%$ (i.e., the threshold we considered for our class 3). Finally, FlexMatcher is not parameterizable with a quality threshold and already provides a binary output (i.e., non-joinable/joinable). In this case, non-joinable maps to uninteresting and joinable to interesting.

\medskip

\begin{figure}[h]
	\begin{center}
		\includegraphics[width=1\linewidth]{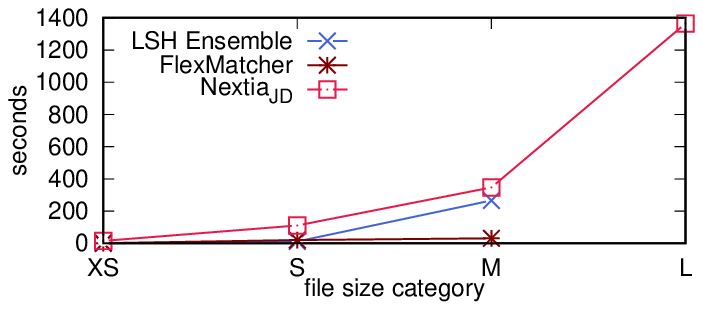}
		\vspace{-2em}
		\caption{Pre runtime}
		\label{fig:prePerformance}
	\end{center}
\end{figure}

\noindent\textbf{Results.}
We evaluated the performance of \name, LSH Ensemble and FlexMatcher on each testbed. Both LSH Ensemble and FlexMatcher suffered from scalability issues and were not able to execute testbed $L$. Figure \ref{fig:prePerformance}, depicts the runtime of the \textit{pre} phase for each testbed. We can observe that the runtime of all systems is in the same orders of magnitude, however our \textit{pre} is larger than the rest. This is mainly due to the fact that, to ensure a fair comparison, we did not set Spark on cluster mode. It is well-known that using Spark on centralized mode adds extra overhead of tasks when generating the required data structures, managing partitions, etc. This is not the case for the other systems, which are provided as standalone programs. Nevertheless, \name~benefits from Spark robustness and it is the only approach, even in centralized mode, capable of dealing with a large-scale testbed. Furthermore, we note that \name~is the only solution able to precompute its \textit{pre} step (except for binary meta-features).

\begin{figure}[!b]
	\begin{center}
		\includegraphics[width=1\linewidth]{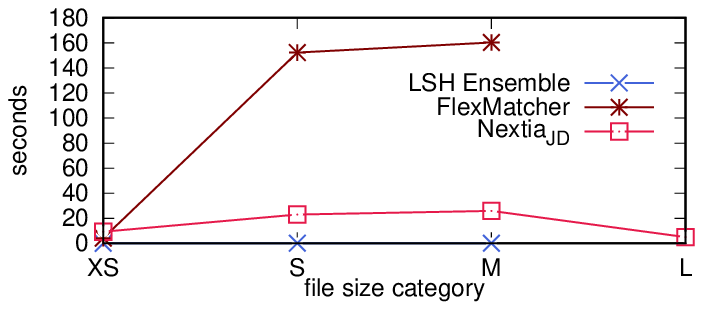}
		\vspace{-2em}
		\caption{Query runtime (\textit{discovery-by-dataset})}
		\label{fig:queryByDatasetPerformance}
	\end{center}
\end{figure}

\begin{figure}[!b]
	\begin{center}
		\includegraphics[width=1\linewidth]{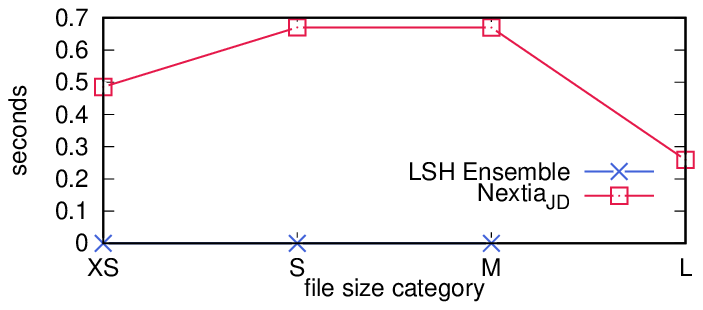}
		\vspace{-2em}
		\caption{Query runtime (\textit{discovery-by-attribute})}
		\label{fig:queryByAttributePerformance}
	\end{center}
\end{figure}

Regarding the computational performance of the \textit{query} phase, we distinguish the \textit{discovery-by-attribute} and \textit{discovery-by-dataset} scenarios, respectively in Figures \ref{fig:queryByDatasetPerformance} and \ref{fig:queryByAttributePerformance}. Note that \textit{discovery-by-attribute} is not available in FlexMatcher. LSH Ensemble excels in both \textit{query} tasks. This is due to the fact there is no mechanism to persist the index and this task is reduced to an in-memory lookup. FlexMatcher also benefits from fully running in memory but, in this case, the \textit{query} step suffers from the need to compute some on-the-fly learning models.
As general observation, both approaches are thought to compute their core structures and run in memory, which is the main reason hindering their ability to scale-up.
Overall, \name~shows a good behaviour in the \textit{query} step. Importantly, \name~is not affected by the dataset cardinality at \textit{query} time. Indeed, the runtime is directly proportional to the number of attributes, or profiles, to compare.

We now put the focus on comparing the predictive performance of the three approaches. Figure \ref{fig:confusionMatricesBinaryNexiavsLSH} and Table \ref{tab:allSystemsAccuracyRecall}, depict, respectively, the confusion matrices and performance metrics using the binary class mapping. Relevantly, the predictive quality of \name~and LSH Ensemble are comparable. While LSH Ensemble finds more true positives, it generates much more false positives. As result, \name~precision is better. Finally, and aligned with our claim that contemporary profile-based data discovery methods fall short in terms of quality, FlexMatcher generates an extremely large number of false positives reducing its overall quality and making it unfeasible for large-scale scenarios.

\begin{figure}[h!]
		\hspace{-1em}
		\begin{subtable}{.4\linewidth}
			\centering
			\caption{\label{tab:nextia-XS-S-M}\hspace{1em} \name}
			\[
			\raisebox{-2\baselineskip}{%
				\begin{tabular}[b]{c@{\hspace*{0.5em}}c} \vspace{-3em}
				& {}
				\\
				\begin{sideways}
				\phantom{..}True
				\end{sideways}
				\begin{sideways}
				\phantom{..}class
				\end{sideways}
				& \raisebox{0.5\height}%
				{\(
					\begin{matrix}
					{} & \phantom{000000} & \phantom{000000} \\
					\textbf{1} & \cellcolor{d4}166 & \cellcolor{d1}915 \\
					\textbf{0} & \cellcolor{d1}419119 & \cellcolor{d4}129 \\
					{} & \textbf{0} & \textbf{1}
					\end{matrix}
					\)%
				} \\
				& Predicted class
				\end{tabular}}
			\]
		\end{subtable}
		\begin{subtable}{.05\linewidth}
			\hspace*{\fill}
		\end{subtable}
		\begin{subtable}{.4\linewidth}
			\caption{\label{tab:lshensamble-XS-S-M}\hspace{1em} LSH Ensemble}
			\centering
			\[ 
			\raisebox{-2\baselineskip}{%
				\begin{tabular}[b]{c@{\hspace*{0.5em}}c} \vspace{-3em}
				& {}
				\\
				\begin{sideways}
				\phantom{..}True
				\end{sideways}
				\begin{sideways}
				\phantom{..}class
				\end{sideways}
				& \raisebox{0.5\height}%
				{\(
					\begin{matrix}
					{} & \phantom{000000} & \phantom{000000} \\
					\textbf{1} & \cellcolor{d4}55 & \cellcolor{d1}1026 \\
					\textbf{0} & \cellcolor{d1}418338 & \cellcolor{d4}910 \\
					{} & \textbf{0} & \textbf{1}
					\end{matrix}
					\)%
				} \\
				& Predicted class
				\end{tabular}}
			\]
		\end{subtable}
		
		\begin{subtable}{.4\linewidth}
			\caption{\label{tab:flexmatcher-XS-S-M}\hspace{1em} FlexMatcher}
			\centering
			\[ 
			\hspace{-0.5em}
			\raisebox{-2\baselineskip}{%
				\begin{tabular}[b]{c@{\hspace*{0.5em}}c} \vspace{-3em}
				& {}
				\\
				\begin{sideways}
				\phantom{..}True
				\end{sideways}
				\begin{sideways}
				\phantom{..}class
				\end{sideways}
				& \raisebox{0.5\height}%
				{\(
					\begin{matrix}
					{} & \phantom{000000} & \phantom{000000} \\
					\textbf{1} & \cellcolor{d4}572 & \cellcolor{d1}509 \\
					\textbf{0} & \cellcolor{d1}381493 & \cellcolor{d4}37755 \\
					{} & \textbf{0} & \textbf{1}
					\end{matrix}
					\)%
				} \\
				& Predicted class
				\end{tabular}}
			\]
		\end{subtable}
	\vspace{-1em}
	\caption{ \label{fig:confusionMatricesBinaryNexiavsLSH} Combined confusion matrices for each system on testbeds $XS,S,M$}
\end{figure}
\begin{table}[h!]
	\centering
	\resizebox{0.4\textwidth}{!}{
		\begin{tabular}{|c|c|c|c|}
			\hline
			& \textbf{Precision} & \textbf{Recall} & \textbf{$\bm{F_1}$ score} \\
			\hline
			\name & $0.8764$ & $0.8464$ & $0.8611$ \\
			\hline
			LSH Ensemble & $0.5299$ & $0.9491$ & $0.6800$ \\
			\hline
			FlexMatcher & $0.0133$ & $0.4708$ & $0.0258$ \\
			\hline
		\end{tabular}
	}
	\caption{\label{tab:allSystemsAccuracyRecall} Performance metrics using binary classes for each system under evaluation on testbeds $XS,S,M$}
\end{table}

Then, Figure \ref{fig:detailedComparisonNextiaLSHEnsamble} drills deeper~ into~ the~ comparison~ between~~ $ $ \name~$ $ and LSH Ensemble. We assigned a quality class to LSH Ensemble by running it several times and using a different containment threshold each time, as defined in our quality classes (i.e., 0.75, 0.5, 0.25 and 0.1). There, we observe relevant differences on the predictions computed. In general, our approach is more conservative, in the sense that we produce less false positives at expenses of sacrificing some true positives. Overall, this improves the precision of our approach by reducing the number of false positives shown to the user. As general observation, both approaches follow slightly different objectives and \name~is more suitable for large-scale scenarios, both for its scale-up capacity and high precision, which guarantees the user will not be overwhelmed with large rankings including false positives.

\begin{figure}[hbtp]
	\begin{center}
		\begin{subtable}{1\linewidth}
			\centering
			\caption{\label{tab:nextia_XS-S-M}\hspace{1em} \name~($XS,S,M$)}
			\[
			\raisebox{-2\baselineskip}{%
				\begin{tabular}[b]{c@{\hspace*{0.5em}}c} \vspace{-3em}
				& {}
				\\
				\begin{sideways}
				\phantom{0}True class
				\end{sideways}
				& \raisebox{0.5\height}%
				{\(
					\begin{matrix}
					{} & \phantom{000000} & \phantom{000000} & \phantom{000000} & \phantom{000000} & \phantom{000000} \\
					\textbf{4} & \cellcolor{d5}3 & \cellcolor{d4}0 & \cellcolor{d3}41 & \cellcolor{d2}253 & \cellcolor{d1}337 \\
					\textbf{3} & \cellcolor{d4}5 & \cellcolor{d3}6 & \cellcolor{d2}111 & \cellcolor{d1}324 & \cellcolor{d2}1 \\
					\textbf{2} & \cellcolor{d3}92 & \cellcolor{d2}3 & \cellcolor{d1}234 & \cellcolor{d2}70 & \cellcolor{d3}1 \\
					\textbf{1} & \cellcolor{d2}4586 & \cellcolor{d1}640 & \cellcolor{d2}257 & \cellcolor{d3}1 & \cellcolor{d4}0 \\
					\textbf{0} & \cellcolor{d1}410433 & \cellcolor{d2}2270 & \cellcolor{d3}604 & \cellcolor{d4}56 & \cellcolor{d5}1 \\
					{} & \textbf{0} & \textbf{1} & \textbf{2} & \textbf{3} & \textbf{4}
					\end{matrix}
					\)%
				} \\
				& Predicted class
				\end{tabular}}
			\]
		\end{subtable}
\\
		\begin{subtable}{1\linewidth}
			\centering
			\caption{\label{tab:LSH_XS-S-M}\hspace{1em} LSH Ensemble ($XS,S,M$)}			
			\[ 
			\raisebox{-2\baselineskip}{%
				\begin{tabular}[b]{c@{\hspace*{0.5em}}c} \vspace{-3em}
				& {}
				\\
				\begin{sideways}
				\phantom{0}True class
				\end{sideways}
				& \raisebox{0.5\height}%
				{\(
					\begin{matrix}
					{} & \phantom{000000} & \phantom{000000} & \phantom{000000} & \phantom{000000} & \phantom{000000} \\
					\textbf{4} & \cellcolor{d5}1 & \cellcolor{d4}0 & \cellcolor{d3}0 & \cellcolor{d2}5 & \cellcolor{d1}628 \\
					\textbf{3} & \cellcolor{d4}0 & \cellcolor{d3}0 & \cellcolor{d2}54 & \cellcolor{d1}84 & \cellcolor{d2}309 \\
					\textbf{2} & \cellcolor{d3}17 & \cellcolor{d2}27 & \cellcolor{d1}180 & \cellcolor{d2}111 & \cellcolor{d3}65 \\
					\textbf{1} & \cellcolor{d2}4606 & \cellcolor{d1}191 & \cellcolor{d2}185 & \cellcolor{d3}497 & \cellcolor{d4}5 \\
					\textbf{0} & \cellcolor{d1}412425 & \cellcolor{d2}447 & \cellcolor{d3}260 & \cellcolor{d4}227 & \cellcolor{d5}5 \\
					{} & \textbf{0} & \textbf{1} & \textbf{2} & \textbf{3} & \textbf{4}
					\end{matrix}
					\)%
				} \\
				& Predicted class
				\end{tabular}}
			\]
		\end{subtable}%
	\end{center}
	\vspace{-1em}
	\caption{Confusion matrices using 5 quality classes for \name~and LSH Ensemble \label{fig:detailedComparisonNextiaLSHEnsamble} }
\end{figure}

\subsection{Scalability}

Since we target large-scale scenarios, we evaluate the scalability of our \name~in such settings. We experiment with different file sizes both in terms of rows and columns.
\medskip

\noindent\textbf{Methodology.}
As shown in Figure \ref{fig:prePerformance}, our most intensive task with regard to computational resources is the generation of attribute profiles from datasets. We hence performed a stress test of this component by means of two experiments. First, we generated a 10GB base CSV file with 5 columns and systematically extended it in batches of 10GBs, up to 60GBs. Next, we followed a similar strategy with regard to columns. We created a 20GB base file that was systematically extended with a new duplicate column each time. The resulting files were stored in a Hadoop HDFS cluster, using the default block size and replication parameters. In order to simulate a realistic large-scale scenario, we also converted each input file to Apache Parquet\footnote{\url{https://parquet.apache.org/}} format. Parquet is an specialized hybrid layout that fragments data into row group partitions (i.e., physically-independent columns), while it also embeds numerous statistics to optimize queries. To evaluate the scalability of our approach in terms of distribution, we compute the profiling runtime using $n$ Spark workers (cf. HDFS datanodes) in the range $1 \ldots 3$.

\medskip

\begin{figure}[h!]
	\begin{center}
		\includegraphics[width=1\linewidth]{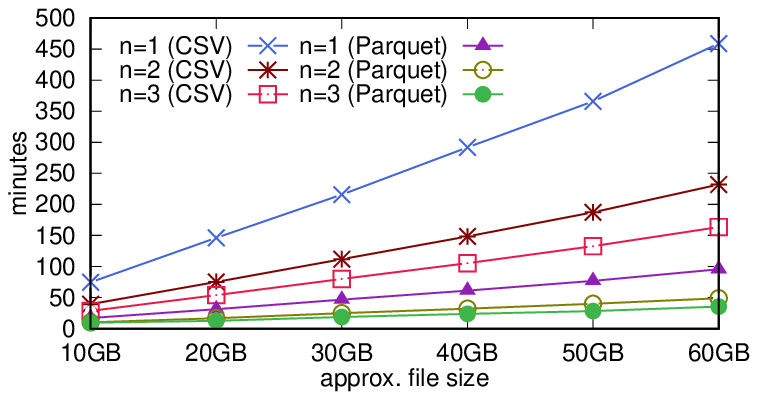}
		\vspace{-2em}
		\caption{Profiling runtime over an increasing file size}
		\label{fig:horizontalScalability}
	\end{center}
\end{figure}

\noindent\textbf{Results.}
Figure \ref{fig:horizontalScalability} depicts the profiling runtime for an increasing file size. Regardless of the number of workers and data format used, the runtime linearly scales with the file size. As expected, profiling Parquet files are much more efficient than CSV ones (i.e., an approximate 4x to 5x speed-up), as we can benefit from statistics and compression when computing certain meta-features. 

\begin{figure}[h!]
	\begin{center}
		\includegraphics[width=1\linewidth]{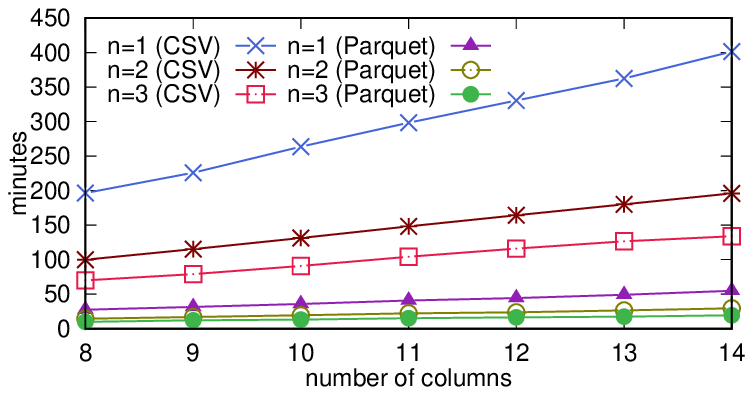}
		\vspace{-2em}
		\caption{Profiling runtime over an increasing number of columns}
		\label{fig:verticalScalability}
	\end{center}
\end{figure}

As depicted in Figure~\ref{fig:verticalScalability}, we can also observe that the profiling runtime trend scales linearly with the number of columns. Similarly to the previous case, using Parquet significantly speeds up the process, here with a 7x to 8x factor. 
Finally, in Table \ref{tab:profileSize}, we show the average profile size per each testbed. The disk usage is proportional to both the number of rows and columns. Although the number of columns is the leading factor for the profiling size, the dataset cardinality impacts on the size of some meta-features (e.g., frequent words, soundex, etc.). In any case, the profile sizes are reasonable. Thus, they can be precomputed offline and stored together with the dataset as metadata. The only exception to this would be binary meta-features. As final conclusion, these experiments show that our approach does not introduce any blocking factor hindering parallelism and can fully benefit from it.

\begin{table}[h!]
	\centering
	\resizebox{0.3\textwidth}{!}{
		\begin{tabular}{|c|c|}
			\hline
			\textbf{Testbed} & \textbf{Average profile size (KBs)} \\
			\hline
			$XS$ & $132$ \\
			\hline
			$S$ & $343$ \\
			\hline
			$M$ & $282$ \\
			\hline
			$L$ & $398.1$ \\
			\hline
		\end{tabular}
	}
	\caption{\label{tab:profileSize} Average profile size per testbed}
\end{table}

\subsection{Discovery of semantic non-syntactic relationships}

This last experiment aims at scrutinizing the behaviour of our model in front of semantic non-syntactic joinable pairs (see Section \ref{sec:background}).

\medskip

\noindent\textbf{Methodology.}
We manually curated a test dataset $D$ of 532 semantic non-syntactic attribute pairs. In order to devise their true label, we applied the following three transformations: (1) string cleaning (e.g., remove accents, dashes, etc.), (2) dictionary lookup (e.g., country codes to country names, author names to full names, etc.), and (3) unwind collections. Accordingly, we were able to generate an approximate ground truth $D^\prime$ applying the quality definition over the transformed dataset. Using that, we were able to evaluate the performance of \name~on $D$ by comparing the predictions to the true labels in $D^\prime$.

\medskip

\noindent\textbf{Results.}
At first, \name~yielded low predictive performance. A detailed analysis revealed that (i) a strict quality definition and (ii) the chain classifier strategy were drastically penalizing this kind of join relationship. These factors were identified as the main reasons for our high precision and low number of false positives. As consequence, we trained a new set of classifiers (using the same ground truth as shown in Section \ref{sec:predictiveModel}), deactivating the chain classifier strategy and keeping the same join quality definition. Figure \ref{fig:comparisonSemanticNonSyntactic} depicts the results in a confusion matrix. This new model classified 48.87\% of the true positives as classes 1-3. Nonetheless, most of these pairs are misclassified to lower quality classes than their true class. Note that non surprisingly, no pair was predicted as quality 4, due to our strict definition of high quality joins, hard to achieve for semantic non-syntactic attributes. A closer look to the predicted pairs showed the complexity of identifying such kind of pairs with our quality join definition and current set of meta-features. Although these results show that our method only works for semantic and/or syntactic relationships, we do believe they are promising.As a matter of fact, the meta-features used in our classifiers (see Table \ref{tab:metadataList}) limit the ability to detect semantic non-syntactic pairs. Specifically, \textit{Syntactic} meta-features are mostly writing off such pairs. Yet, we were able to identify some due to the role played by the \textit{Cardinalties} and \textit{Value distribution} meta-features. We conclude that a new model should be devised, with specific meta-features to capture the special nature of these relationships.

\begin{figure}[htbp!]
	\[ 
	\raisebox{-2\baselineskip}{%
		\begin{tabular}[b]{c@{\hspace*{0.5em}}c} \vspace{-3em}
		& {}
		\\
		\begin{sideways}
		\phantom{0}True class
		\end{sideways}
		& \raisebox{0.5\height}%
		{\(
			\begin{matrix}
			{} & \phantom{000000} & \phantom{000000} & \phantom{000000} & \phantom{000000} & \phantom{000000} \\
			\textbf{4} & \cellcolor{d5}90 & \cellcolor{d4}46 & \cellcolor{d3}47 & \cellcolor{d2}22 & \cellcolor{d1}0 \\
			\textbf{3} & \cellcolor{d4}36 & \cellcolor{d3}24 & \cellcolor{d2}32 & \cellcolor{d1}7 & \cellcolor{d2}0 \\
			\textbf{2} & \cellcolor{d3}37 & \cellcolor{d2}15 & \cellcolor{d1}20 & \cellcolor{d2}0 & \cellcolor{d3}0 \\
			\textbf{1} & \cellcolor{d2}68 & \cellcolor{d1}23 & \cellcolor{d2}3 & \cellcolor{d3}1 & \cellcolor{d4}0 \\
			\textbf{0} & \cellcolor{d1}41 & \cellcolor{d2}19 & \cellcolor{d3}1 & \cellcolor{d4}0 & \cellcolor{d5}0 \\
			{} & \textbf{0} & \textbf{1} & \textbf{2} & \textbf{3} & \textbf{4}
			\end{matrix}
			\)%
		} \\
		& Predicted class
		\end{tabular}}
	\]
	\vspace{-1em}	\caption{ \label{fig:comparisonSemanticNonSyntactic} Semantic non-syntactic attributes identified}
\end{figure}

\section{Conclusions and future work}\label{sec:future}
We have presented a novel learning-based approach for data discovery on large-scale repositories of heterogeneous, independently created datasets. Our work is motivated by (i) the poor predictive performance of current profile-based solutions, and (ii) the inability to scale-up of hash-based approaches, as well as their low precision, which is undesirable for large-scale scenarios. In order to overcome these limitations, we propose a scalable method yielding good precision, and grounded on a novel qualitative definition of join quality. Further, we implemented our approach in a tool called \name. We have experimentally shown that despite being a profile-based approach, \name~presents a similar predictive performance to that of hash-based solutions, yet better adapted for large-scale scenarios, while benefiting from linear scalability.

As future work, we look for adapting our approach to detect semantic non-syntactic join relationships (i.e., requiring some simple transformation on the values before joining). Based on such predictions, the system should be able to propose the required transformations to join. 

\section*{Acknowledgments}

This work is partly supported by the GENESIS project, funded by the Spanish Ministerio de Ciencia e Innovaci\'{o}n grant TIN2016-79269-R. Javier Flores is supported by contract 2020-DI-027 of the Industrial Doctorate Program of the Government of Catalonia.

\bibliographystyle{ACM-Reference-Format}
\bibliography{main}

\end{document}